\documentclass[]{interact}

\usepackage{epstopdf}
\usepackage[caption=false]{subfig}
\usepackage{appendix}
\usepackage{tikz} 
\usetikzlibrary{calc, angles, quotes}
\usepackage{xcolor}

\usepackage[numbers,sort&compress,merge]{natbib}
\bibpunct[, ]{[}{]}{,}{n}{,}{,}
\renewcommand\bibfont{\fontsize{10}{12}\selectfont}

\theoremstyle{plain}

\theoremstyle{definition}

\theoremstyle{remark}

\usepackage[numbers,sort&compress,merge]{natbib}
\bibpunct[, ]{[}{]}{,}{n}{,}{,}
\renewcommand\bibfont{\fontsize{10}{12}\selectfont}

\begin{document}


\title{Vibrational, non-adiabatic and isotopic effects
  in the dynamics of the H$_2$ + H$_2^+$ $\rightarrow$ H$_3^+$ + H reaction:
  application to plasma modeling}

\author{
  \name{
      P. del Mazo-Sevillano\textsuperscript{a,b},
      D. F{\'e}lix-Gonz{\'a}lez\textsuperscript{b},
      A. Aguado\textsuperscript{b},
      C. Sanz-Sanz\textsuperscript{b},
      D.-H. Kwon\textsuperscript{c}, 
      O. Roncero\textsuperscript{d}\thanks{CONTACT O. Roncero. Email: octavio.roncero@csic.es}
   }
   \affil{
   \textsuperscript{a} FU Berlin, Department of Mathematics and Computer Science, Arnimallee 12, 14195 Berlin, Germany\\
      \textsuperscript{b} Unidad Asociada UAM-IFF-CSIC,
          Departamento de Qu{\'\i}mica F{\'\i}sica Aplicada, Facultad de
          Ciencias M-14, Universidad Aut{\'o}noma de Madrid, 28049, Madrid, Spain          \\
      \textsuperscript{c}  Nuclear Physics Application Research Division, Korea Atomic Energy Research Institute, Daejeon 34057, Republic of Korea\\
      \textsuperscript{d} Instituto de F{\'\i}sica Fundamental, IFF-CSIC, c/ Serrano 123, 28006 Madrid, Spain
   }
}

\maketitle

\begin{abstract}
  The title reaction is studied  using a quasi-classical trajectory method for collision energies
  between 0.1 meV and 10 eV,  considering the vibrational excitation of H$_2^+$ reactant.
  A new potential energy surface is developed based on a Neural Network many body correction of 
  a triatomics-in-molecules potential, which significantly improves the accuracy of
  the potential up to energies of 17 eV, higher than in other previous fits.
  The effect of the fit accuracy and the non-adiabatic transitions on the dynamics are analyzed in detail.
  The reaction cross section for collision energies above 1 eV increases significantly with the increasing of the vibrational
  excitation of H$_2^+$($v'$), for values up to $v'$=6. The total reaction cross section (including
  the double fragmentation channel) obtained for $v'$=6 matches the new
  experimental results obtained by Savic, Schlemmer and Gerlich \cite{Savic-etal:20}.
  The differences among several experimental setups, for collision energies above 1 eV,
  showing cross sections scattered/dispersed over a rather wide interval, can be explained by the differences in the
  vibrational excitations obtained in the formation of H$_2^+$ reactants.  On the contrary, for collision energies below 1 eV, the cross section
 is determined by the long range behavior of the potential and do not depend strongly on the vibrational state of H$_2^+$.
 In addition in this study, the calculated reaction cross sections are used in a plasma model
 and compared with previous results.  We conclude that the efficiency of the formation of H$_3^+$ in the plasma is affected by the potential energy surface used. 

\end{abstract}


\section{Introduction}

Hydrogen, as the most abundant element in the Universe, plays a fundamental role
in  star formation and the chemical evolution of molecular Universe.
Its molecular forms are H$_2$, H$_2^+$ and H$_3^+$ \cite{Oka:13}. In evolved galaxies,
the formation of H$_2$ is usually attributed to atomic hydrogen recombination
on cosmic  grains and ices \cite{Glover:03,Herbst-Millar:08,Wakelam-etal:17}.
H$_2^+$ is rapidly formed by cosmic rays or electrons,
and it collides with H$_2$ to form H$_3^+$ in the reaction
\begin{eqnarray}\label{H3+reaction}
  {\rm H}_2(v,j) +  {\rm H}_2^+(v',j') \rightarrow {\rm H}_3^+ + {\rm H}.
\end{eqnarray}
 Once the molecular forms of hydrogen are formed, the 
chemistry in space starts with the formation of the first hydrides.
In cold clouds, the most abundant ion is  H$_3^+$, which is considered to be the
universal protonator \cite{Millar-etal:89,Pagani-etal:92,Tennyson:95,McCall-Oka:00}
through the proton hop reaction \cite{Oka:12,Oka:13}
\begin{eqnarray}\label{proton-hop-reaction}
   {{\rm H}}_3^+  + {\rm M} \rightarrow {{\rm HM}}^+ + {\rm H}_2,
\end{eqnarray}
where M is an atom or a molecule.
The HM$^+$ cations are very reactive and trigger many
chemical networks giving rise to most of the molecular systems
detected in space \cite{Watson:73,Herbst-Klemperer:73,Millar-etal:89,Pagani-etal:92,Tennyson:95,McCall-Oka:00}.
In cold environments,  $M$ colliders in Eq.~(\ref{proton-hop-reaction}) deposit
on ices, and H$_3^+$  reacts only with the most abundant molecule, H$_2$, as 
\begin{eqnarray}\label{H3+-exchange-reaction}
  {\rm H}_2 + {\rm H}_3^+ \rightarrow {\rm H}_3^+ + {\rm H}_2,
\end{eqnarray}
a proton exchange reaction, which is constrained  by nuclear spin
statistic \cite{Quack:77,Uy-etal:97,Oka:04,Gerlich-etal:06,Park-Light:07b,Hugo-etal:09,Crabtree-etal:11b},
and is the responsible for the ortho/para ratio of H$_3^+$.
When the collider is HD, reaction~(\ref{H3+-exchange-reaction}) is the responsible of
H$_3^+$ deuteration \cite{Smith-etal:82,Giles-etal:92,Uy-etal:97,Cordonnier-etal:00,%
  Gerlich-Schlemmer:02,Gerlich-etal:02,Hugo-etal:09,Crabtree-etal:11}. The deuterated
species formed following Eq.~(\ref{proton-hop-reaction}) produce the observed
high relative abundance of deuterated species, estimated as $\approx$ 10$^4$ times
higher than the D/H ratio of the galaxy \cite{Millar:02,Pagani-etal:09,Bell-etal:11,Fontani-etal:11}. 
This high deuteration efficiency  is attributed to the 
zero-point energy differences among the H$_3^+$ deuterated species,
very significant at the low temperatures of cold molecular clouds
\cite{Gerlich:90,Gerlich-Schlemmer:02}.

Hydrogen plasma \cite{Mendez-etal:06,Jimenez-Redondo-etal:11,Tanarro-Herrero:11}, apart from their technological applications in industry, medicine
and  fusion reactors \cite{Grill:94,Lieberman-Lichtenberg:05}, can be considered as a prototype
for Early Universe models \cite{Glover:03,Savin-etal:04,Coppola-etal:11,Indriolo-McCall:12,Coppola-etal:13}.
In the absence of other species but hydrogen, 
the molecular species are simply H$_2$, H$_2^+$ and H$_3^+$, but H$^+$   also exists.
The distribution of the ions affects the determination of hydrogen particle flux in the
plasma \cite{Chai, Perillo}. The reaction in Eq.~(\ref{H3+reaction}) is an important process for modeling
H$_2$ plasma
at low electron and molecular temperatures ($T_e \sim 5$~eV, $T_m  \sim 0.1$~eV)
since the process is the only dominant process for the formation of H$_3^+$ \cite{Lishev}.
Therefore, an accurate cross section is essential
for the determination of the density of H$_3^+$ in plasma models.
Moreover, the rovibrationally resolved cross sections are required for collisional-radiative (CR) spectroscopic
modeling of molecular hydrogen which can be applied to a fusion detached plasma \cite{Sawada, Dirk}.
The isotopic effect on the reaction is also essential
for plasma modeling in nuclear fusion tokamak \cite{Verhaegh}.

The reaction in Eq.~(\ref{H3+reaction}) has been widely studied
experimentally \cite{Giese-Maier:63,Neynaber-Trujillo:68,Gentry-etal:75,Specht-etal.75,Krenos-etal:76,Koyano-Tanaka:80,Anderson-etal:81,%
   Shao-Ng:86,Pollard-etal:91,Glenewinkel-Meyer-Gerlich:97}
 and  theoretically \cite{Krenos-etal:76,Eaker-Schatz:85,Badenhoop-etal:87,Stine-Muckerman:78,Baer-Ng:90,Alijah-Varandas:08,Sanz-Sanz-etal:15}.
 In the  1 meV and 1 eV energy interval,
 an excellent agreement between experimental \cite{Glenewinkel-Meyer-Gerlich:97}
 and theoretical \cite{Sanz-Sanz-etal:15} reactive cross sections has been achieved.
 Recently, this reaction has been studied in two extreme regimes,
 at ultra cold energies \cite{Allmendinger-etal:16,Allmendinger-etal:16b,Hoveler-tal:21,Merkt-etal:22}  
 and at high collision energies \cite{Savic-etal:20} up to 10 eV. At ultra cold energies, many
 isotopic variants have been studied, and the reactive cross section shows a Langevin-like behavior so that
 the measured reactive cross sections, in relative units, shows also excellent agreement with
 the theoretical simulations \cite{Sanz-Sanz-etal:15}. However, recent experiments performed
 at higher  collision energies (1-10 eV) \cite{Savic-etal:20}
 differ considerably from previous theoretical simulations \cite{Krenos-etal:76,Eaker-Schatz:85,Sanz-Sanz-etal:15}.
 Moreover, for collision energies above 1 eV,
 the new experimental measurements show differences with previous experimental
 ones \cite{Giese-Maier:63,Neynaber-Trujillo:68,Gentry-etal:75,Specht-etal.75,Shao-Ng:86,Pollard-etal:91},
 all of them scattered in a rather
 wide interval of cross sections.

 There are two main goals in this work. First, we focus on
 the theoretical simulations of  the H$_3^+$  formation
 reaction, Eq.~(\ref{H3+reaction}), to reproduce the new
 experimental data above 1 eV \cite{Savic-etal:20}. Second, we study 
 hydrogen or deuterium plasma models in order to analyze the effects of the calculated cross sections
 and rates on the H$_3^+$ or D$_3^+$ densities.
 
 This work is organized as follows. In section 2,  we develop new potential energy surfaces (PESs),
 which include  non-adiabatic effects and increase the accuracy of the fit. This new fit
 uses a Neural Network (NN) method to describe the four-body term, to improve a zero-order
 description using a Triatomics-in-Molecule treatment, which accurately
 fits  very precise {\it ab initio}
 calculations, over a  broader  energy interval than previous fits,
 up to 17 eV. In section 3,
 we study the reaction dynamics
 using a quasi-classical trajectory (QCT) method, including transitions among different electronic
 states, using the fewest switches method of Tully \cite{Tully:90}. The reaction
 dynamics is studied for collision energies between 1 meV and 10 eV, and for several vibrational states
 of H$_2^+$ reactant, as well as for the deuterated reaction,
 focusing on high energy reactive cross sections recently measured by Savic {\it et al.} \cite{Savic-etal:20}.
 In section 4, we investigate how the calculated reactive cross sections affect  the
 population density of H$_3^+$ and D$^+_3$ in a CR  models of H$_2$/D$_2$ plasma.
 Finally, in section 5 some conclusions
 are extracted.

\section{Potential energy surfaces of H$_4^+$ }

In this work several PESs are used, and are listed here to clarify the differences:
\begin{itemize}
    \item {\bf PES1}: This PES developed by Sanz-Sanz {\it et al.} \cite{Sanz-Sanz-etal:13}, is the most
    accurate developed so far for this system. It is built as a sum
    of a triatomics-in-molecules (TRIM) term, $H^{TRIM}$, plus a four body term, $H^{MB}$.
    The TRIM term is a generalization of
    the Diatomics-in-Molecules (DIM) \cite{Ellison:63,Ellison-etal:63,Tully:80,Kuntz:79} method,
    in which the electronic Hamiltonian is factorized as a sum
    of triatomic and diatomic fragments as \cite{Sanz-Sanz-etal:13,Aguado-etal:10}
    \begin{eqnarray}
     \hat{H}_e^i = \sum_{n>i,o>n} \hat{H}_{ino}^+ (n-i,o-i) - \sum_{p>i} \hat{H}_{ip}^+(p-i)
    \end{eqnarray}
    where $\hat{H}_{ip}^+$ are the monoelectronic Hamiltonians of H$_2^+$ fragments and
    $\hat{H}_{ino}^+(n-i,o-i)$ are the bielectronic Hamiltonians (for $n-i,o-i$ electrons) describing
    the H$_3^+$ system for the $ino$ nuclei. The TRIM representation consists of a 8$\times$8 matrix, whose
    elements have H$_3^+$ and H$_2^+$ matrix elements of different electronic configurations, in which
    each hydrogen atom is described by a 1s function, except one corresponding to H$^+$.
    In PES1 \cite{Sanz-Sanz-etal:13},
    the triatomic terms included in the TRIM matrix were built
    as the 3$\times$3 DIM matrix for each H$_3^+$ fragment plus a three-body
    term added to describe the ground state of either singlet or triplet symmetry. The non-adiabatic terms
    in these triatomic fragments are therefore approximated by those obtained in a DIM treatment.

    PES1 is built  adding a four-body correction (MB term) to this TRIM description. This MB term is expressed
    as a linear combination of four-body polynomials that are symmetric
    with respect to the permutations of identical nuclei (Permutationally invariant polynomials or PIP). These PIP
    are built in terms of Rydberg polynomials of the interatomic distances 
    ($p_i = r_i \exp(-\alpha_i r_i)$) \cite{Aguado-Paniagua:92, Aguado-etal:94,Tablero-etal:99}.
    The set of linear and non linear coefficients, $\alpha_i$, are optimized  to 
    minimize the difference with the {\it ab initio} energies obtained with the multi reference configuration interaction (MRCI) method
    using the aug-cc-pV5Z basis set \cite{Dunning:89}. In PES1, the
    same four-body correction term is added to all the diagonal elements of the TRIM  matrix.

    The reactive cross sections calculated on this PES1 shows an excellent agreement with the experimental
    results for collision energies between 1 meV and 1 eV
    \cite{Sanz-Sanz-etal:15, Savic-etal:20, Allmendinger-etal:16, Allmendinger-etal:16b}.

    \item {\bf PESTRIM8$\times$8 and PESTRIM1$\times$1}: This PES is based on the TRIM model explained 
      above, and is an improvement made in this work. The main difference is that the triatomic H$_3^+$
      is represented by 3$\times$3 diabatic 
    matrices fitting its three lowest singlet and triplet electronic states \cite{Aguado-etal:21}
    to MRCI {\it ab initio} energies extrapolated to the complete basis set (CBS) limit \cite{Velilla-etal:10}.
    This improvement is crucial to study the 
    non-adiabatic dynamics of H$_4^+$. Analytical derivatives of the potential energy surfaces and 
    non-adiabatic couplings are calculated based on the Hellmann-Feynmann theorem
    as described in \cite{Sanz-Sanz-etal:15}.
    The full PESTRIM8$\times$8 potential correspond to the eight adiabatic eigenvalues including the
    non-adiabatic coupling terms. PESTRIM1$\times$1 only considers the ground adiabatic energy.

    \item {\bf PES-NN}: PESTRIM1$\times$1 lacks high accuracy in the interaction region, i.e. where H$_4^+$
    is formed, while showing excellent agreement for H$_3^+$ + H or H$_2$ + H$_2^+$ rearrangement channels.
    For this reason a four-body neural network term is added to PESTRIM1$\times$1 to improve the accuracy of the
    system, as described below.
\end{itemize}

    \subsection{Many Body Neural Network term}

Many body potential energy terms are widely used to represent potential energy surface of chemical 
systems, either in a many body expansion \cite{Sorbie:75, Murrell:76, Aguado-etal:94},
where one up to $N$-body terms are summed, or as correction
terms of a zero-order description of the potential,  described
by the TRIM method \cite{Aguado-etal:10,Sanz-Sanz-etal:13} or by a reactive force field matrix
\cite{Zanchet-etal:18, Roncero-etal:18, Li-etal:20}.

In this work a many body neural network (MB-NN) \cite{Li-etal:20} potential energy term
has been built as a PIP-NN \cite{Jiang-etal:13}, in which the neural network is fed with
a permutational invariant polynomial representation of the molecular geometry. 
A PIP is constructed by projecting a polynomial of the interatomic distances into the totally symmetric
irreducible representation of the desired permutation group. A generator of these  PIPs is defined as:
\begin{equation}
    P_n = {\hat S}\prod_{i=1}^{N_d} p_i^{l_i^n}(r_i)
    \label{eq:polynomial_descriptor}
\end{equation}
where $N_d$ is the number of interatomic distances,
$p_{i}$ is a function of the interatomic distance $r_i$,
$l_i^n$ is the exponent of the $i$th monomial for the $n$th polynomial
and $ {\hat S}$ is the projector to the totally symmetric irreducible representation
of the permutation group.
A common choice of $p(r)$ in PIP-NN-PES is
the decaying exponential $p_i(r) = \exp(-\alpha_i r)$.

The set of PIP has to be carefully filtered so that it 
is purely composed of $N$-body functions. This means that any of these functions evaluated on a geometry
where the $N$ bodies are not closely interacting should be zero. In case these terms included any lower
body functions, we have shown that it would add spurious interactions
between fragments \cite{delMazoSevillano-etal:21}, which specially
affects the long range regions.

In appendix
\ref{ap:PIP} it is shown that a polynomial in Eq. \eqref{eq:polynomial_descriptor} can be represented 
as a graph, and that the subset of $N$-body 
polynomials corresponds to those whose respective graph is connected, meaning that there exists a path which 
connects all the vertices (particles) of the graph. In this way we can guarantee that any $N$-body 
polynomial, with $p(r)$ defined as a decaying exponential or Rydberg function, will tend to zero as any particle
or set of particles moves away from the rest. This will automatically make zero a PIP-PES and
provide constant descriptor for a NN-PIP-PES, that returns a constant energy out of the $N$-body region,
which can be trained to be as close to zero as possible and that produces no net force, since its 
derivative with respect to the atom coordinates is zero.

The four-body term developed here is expressed as a feed forward neural net with 11 input
neurons and two hidden layers with $N_2$ = 32 and $N_3$ = 77, and sigmoid non linearities
$\sigma = 1 / (1 + \exp(-x))$:
    
    \begin{equation}
        H^{MB} = b_1^{(3)} + \sum_i^{N_3} \left( w_{1i}^{(3)} \sigma \left( b_i^{(2)} + \sum_j^{N_2} \left( w_{ij}^{(2)} \sigma \left( b_k^{(1)} + \sum_k^{N_1} \left( w_{jk}^{(1)} PIP_k \right) \right)\right)\right)\right)
        \label{eq:MB-NN}
      \end{equation}
      {
      were $\mathbf{w}^{(l)}$ matrix (with elements $ w_{ki}^{(l)}$) and $\mathbf{b}^{(l)}$
      vector (with elements $ b_i^{(l)}$)
}
      represent the trainable weights and bias on layer $l$.
The 11 input neurons correspond to four-body PIPs produced by setting a maximum
polynomial degree max($\sum l_i$) = 5 and maximum monomial degree max($l_i$)=2.      
   All lower body polynomials
    that would introduce spurious energy contributions in reactants and products channels are filtered 
    following the steps detailed in appendix \ref{ap:PIP}.

    The four-body term is trained with NeuralPES, an in-house Python code based on PyTorch \cite{NEURIPS2019_9015}. 
New {\it ab initio} points have been used in the fit of this work,
of higher accuracy of those used in PES1 \cite{Sanz-Sanz-etal:13}.
The energies are obtained using a two point extrapolation method to complete basis
set (CBS) \cite{Velilla-etal:10}, using  the results obtained with
the aug-cc-pV5Z and aug-cc-pV6Z basis sets.
Around 33000 {\it ab initio} points were calculated, including the geometries of Ref. \cite{Sanz-Sanz-etal:13}
and new ones selected to increase the accuracy of the PES above 2 eV. These last new points were
chosen from QCT trajectories at different collision energies (from 1 eV to 10 eV)
taken on the TRIMPES1$\times$1 to populate physically accessible configurations at this high energy regions.
The complete set of {\it ab initio} points
is randomly split into a training set (containing 80\% of the data) and validation and test sets
(with 10\% of the points each). 
The training set
consists on 27294 geometries, with energies up to 17 eV over
the H$_2$ + H$_2^+$ asymptote, mostly corresponding to four-body
interactions, and elongations to lower body geometries.
The training process aims to minimize the root mean squared error between the \textit{ab initio}
 and PESTRIM1$\times$1 + $H^{MB}$ energies.

\subsection{Analysis of the different PESs}

As all the four-body terms described above vanish as the systems tend to reactants or product 
asymptotes, the long-range interactions are purely described by the triatomic terms of the TRIM model. 
In all the triatomic fits considered here, the long range terms for H$_2$ + H$^+$ and H$_2^+$ + H fragments,
for either singlet or triplet states are very precisely described \cite{Aguado-etal:21,Velilla-etal:08,Aguado-etal:00}.
This produces highly accurate
 long-range interaction in the H$_2$ + H$_2^+$ channel \cite{Sanz-Sanz-etal:15}. Allmendinger {\it et al.}
\cite{Allmendinger-etal:16,Allmendinger-etal:16b} applied a new experimental set up to study
H$_2^+$ + H$_2$ $\rightarrow$ H$_3^+$ + H reaction at low temperatures.
The measured cross section was then  scaled to reproduce the cross
section calculated with the PES1 potential \cite{Sanz-Sanz-etal:15} at a single collision energy.
The excellent agreement between calculated and  scaled experimental cross sections
between 0.5 and 5 meV demonstrates
the good behaviour of the long-range interactions included in PES1.
The new PESs introduced in this work, describes the long-range interaction
even more accurately, by using improved long-range interactions in the triatomic
fragments \cite{Aguado-etal:21}.

\begin{table}[]
    \centering
    \begin{tabular}{c|c|c|c|c}
        E$_{c}$ (eV)  &  Points & PES1 (meV) & PESTRIM1$\times$1 (meV) & PES-NN (meV) \\ \hline
        0.0 &       9314 &    16.27 &      145.49 &   10.02 \\\hline
        2.5 &     30856 &   25.08 &      131.21 &   10.41 \\\hline
        5.0 &       32314 &   98.58 &      151.64 &   11.65 \\\hline
        7.5 &     32544 &   334.96 &     161.20 &   12.15 \\\hline
        10.0 &      32622 &   850.30 &     164.88 &   12.37 \\\hline
        12.5 &    33007 &   1390.21 &    170.04 &   12.77 \\\hline
        15.0 &      33189 &   2041.10 &    183.78 &   13.24 \\\hline
    \end{tabular}
    \caption{Root mean squared errors of the ground electronic state for different energy regions,
      defined for $E < E_c$. The 
      number of points on which the RMSE is calculated is presented. The zero of energy is set
    at H$_2$ + H$_2^+$ asymptote, with the two fragment in their equilibrium configuration. }
    \label{tab:total_RMSE}
\end{table}

The RMS errors for the different potential energy surfaces are presented in Table \ref{tab:total_RMSE}, in different
energy intervals. The 
improvement of PES-NN over PESTRIM1$\times$1 is clear in all energy ranges due to the enhancement of the
H$_4^+$ channels as presented in Table \ref{tab:channel_RMSE}. PES1 and PES-NN show comparable RMSE for 
energies below 2 eV, but when {\it ab initio} points above 2 eV are added, PES-NN shows a much higher
accuracy.

\begin{figure}[t]
    \centering
    \includegraphics[scale=0.4]{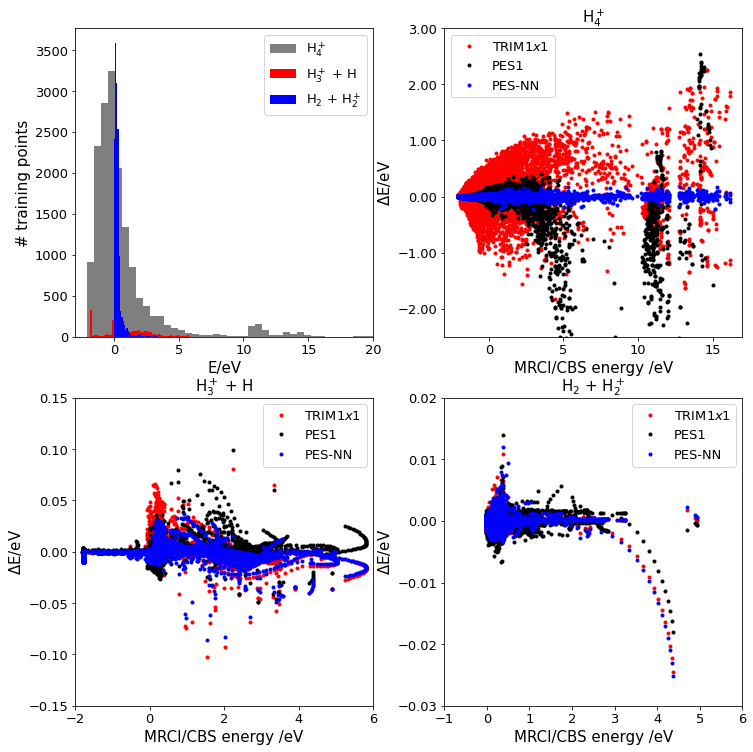}
    \caption{The top left panel shows the energy distribution of training data in three
      regions of the configuration space, as described in the text.
       Top right and bottom panels show the energy difference between the three
    PES considered in this work and {\it ab initio} calculations.}
    \label{fig:trainset_enerdiff}
\end{figure}

The whole configuration space is divided as follows: when all the interatomic distances are below $R_{thres}= 4$ \AA,
      the system is taken to be in H$_4^+$ region;
       if all interatomic distances of a triad of hydrogens are shorter than  $R_{thres}$,
      the region is taken to be H$_3^+$ + H;
      if any two pairs of atoms present interatomic distances shorter than $R_{thres}$, the region
      is denoted as H$_2$ + H$_2^+$; if only one internuclear distance is $<R_{thres}$, the region is called
      H$_2$ + H + H$^+$; otherwise, if all the interatomic distances are large, the system is in fully dissociated.
      Following this division, the top left panel of Figure \ref{fig:trainset_enerdiff} shows the distribution of the data set among the different
      regions as defined above, as a function
of the energy, taking the H$_2$ + H$_2^+$ reactants asymptote as the zero of energy. Most
of the data correspond to H$_4^+$ and geometries
which connect this channel to H$_2$ + H$_2^+$ and H$_3^+$ + H. As can be seen in the top right and bottom
panels, the PESTRIM1$\times$1 is highly accurate in the latter two channels, but shows
deviations in the H$_4^+$ region, up to several eV.  The effect of the four-body
term on PES-NN is decisive for the proper description of H$_4^+$ region. 
PES1 was fitted for energies up to $\approx$ 2 eV,
and it presents large energy deviations over this threshold. On the contrary, PES-NN  
keeps a high precision at high energies, which are the interest of the 
present work.

\begin{table}[b]
    \centering
    \begin{tabular}{c|c|c|c|c}
        Region &  Points & PES1 (meV) & PESTRIM1$\times$1 (meV) & PES-NN (meV) \\ \hline
         H$_4^+$ &      16034 &   3451.60 &   270.27 &   18.97 \\ \hline
         H$_3^+$ + H &     3096 &    9.81 &      12.69 &    8.99 \\ \hline
         H$_2$ + H$_2^+$ &    14031 &   0.87 &      0.74 &     0.82  \\\hline
    \end{tabular}
    \caption{Root mean squared errors of the ground electronic state for different channels, calculated 
    on all the geometries with energy lower than 17 eV.}
    \label{tab:channel_RMSE}
\end{table}

\begin{figure}[t]
  \includegraphics[scale=0.4]{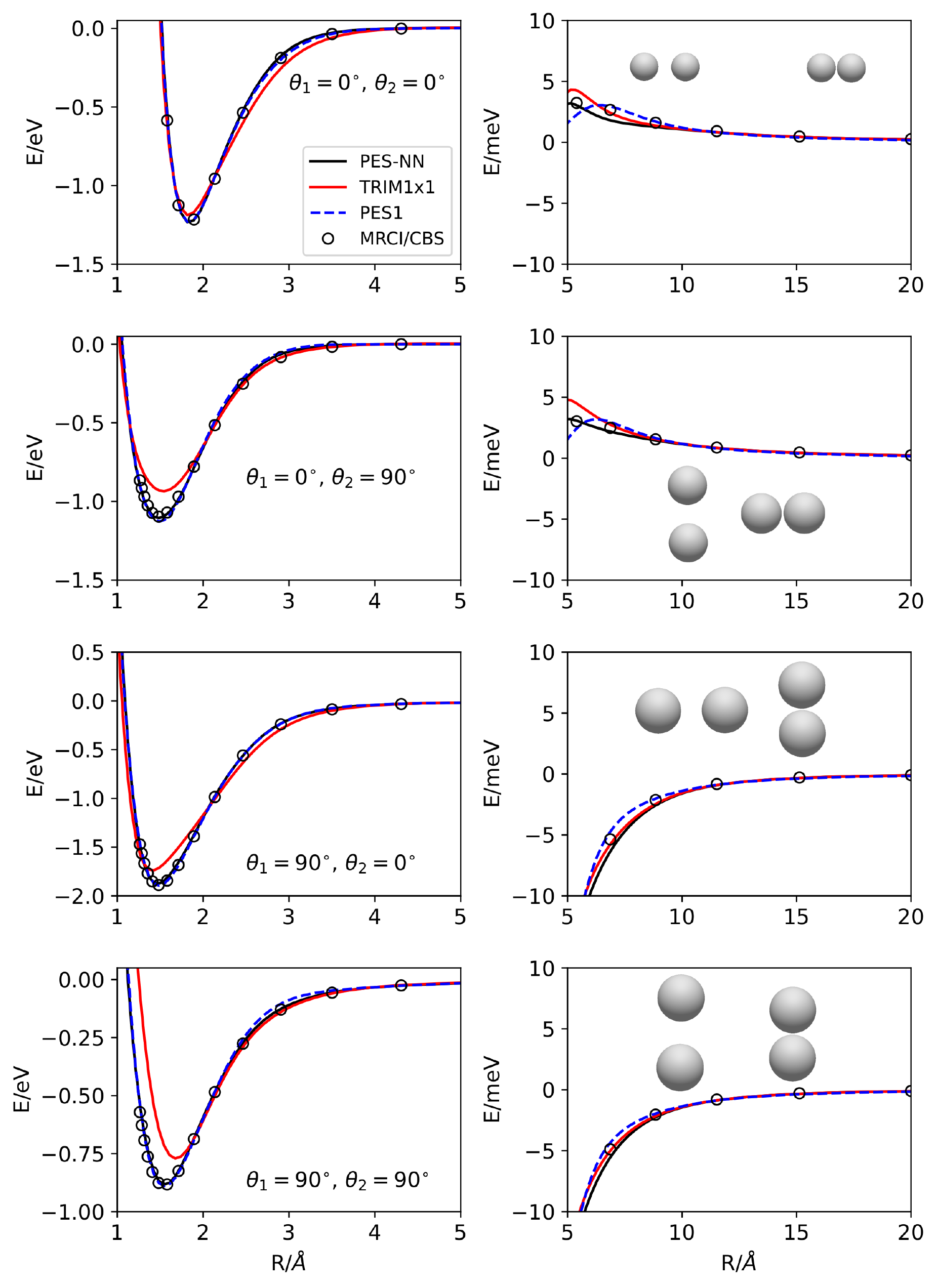}

  \vspace*{-9cm}
  \hspace*{8cm}
    \begin{tikzpicture}[scale=0.9,node distance={0.5mm}, main/.style = {draw, circle}] 
    \node[main] (1) at (0,0) {H};
    \node[main] (2) at (2,1.5) {H};
    \node[main] (3) at (0,-3) {H};
    \node[main] (4) at (2,-3.5) {H};
    \node (a) at ($(1)!0.5!(2)$) {};
    \node (b) at ($(3)!0.5!(4)$) {};
    \draw (1) -- (2) node[above,midway] {$r_2$};
    \draw (3) -- (4) node[below,midway] {$r_1$};
    \draw (a.center) -- (b.center) node[right,midway] {$R$};
    \draw pic[draw,angle radius=0.5cm,"$\theta_2$" shift={(5mm,1mm)}] {angle=b--a--2};
    \draw pic[draw,angle radius=0.5cm,"$\theta_1$" shift={(5mm,2mm)}] {angle=4--b--a};
  \end{tikzpicture}

  \vspace*{3.5cm}
    \caption{H$_2$ + H$_2^+$ approaches for several $\theta_1$ and $\theta_2$ angles and
      $r_1 = 0.740$ \AA\ (for H$_2$) and $r_2 = 1.055$ \AA\ (for H$_2^+$)  the equilibrium distances of both species.
    The inset in the right hand side shows the coordinates used.}
    \label{fig:approach_eq}
\end{figure}

In Figure~\ref{fig:approach_eq}, H$_2$ + H$_2^+$ approaches for different $\theta_1$ and $\theta_2$ angles
in their equilibrium geometry are shown for the three potential energy
surfaces and compared with {\it ab initio} calculations. PESTRIM1$\times$1 tends to 
predict a larger energy for H$_4^+$ geometries, while PES1 and PES-NN yield effectively
the same description. As the interfragment distance increases towards H$_2$ + H$_2^+$ the
four-body terms in both PES1 and PES-NN go to zero and all that remains are the respective
TRIM terms.
The improvement of the new PES-NN is better seen when the diatomic fragments are not at the equilibrium geometry, for
energies above 2 eV.

\begin{figure}[t]
    \centering
    \includegraphics[scale=0.4]{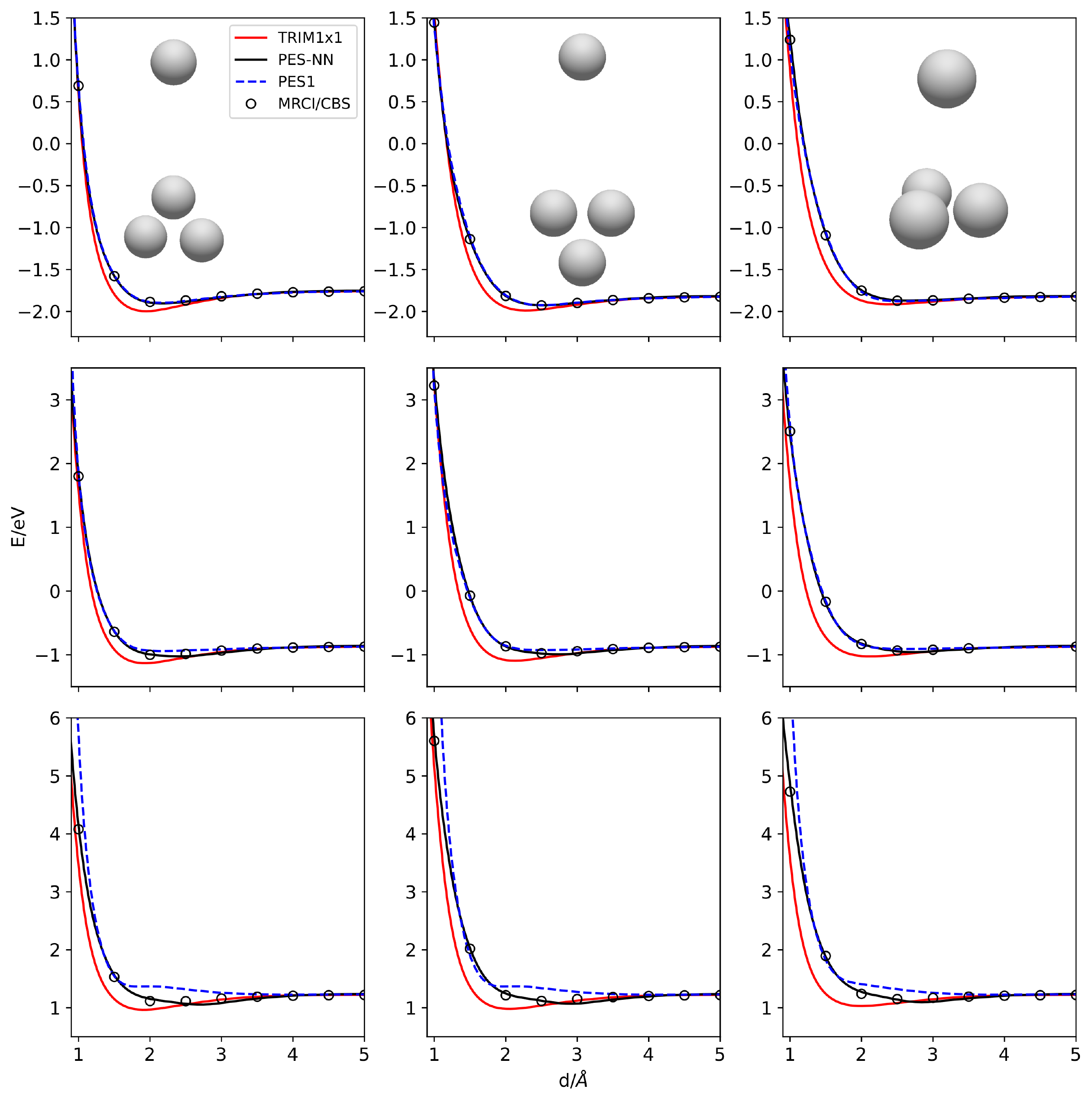}
    \caption{H$_3^+$ + H approaches, as a function of d, the distance of H to the H$_3^+$ center-of-mass.
      The orientations are preserved in the columns. The rows correspond
      to different equilateral H$_3^+$ bond distances: top panels 0.85\AA\ (equilibrium), middle panels to $0.7$ \AA\
      and bottom panels to $0.6$ \AA, respectively.
      { In the top panel, corresponding to H$_3^+$ equilibrium configuration,
        the asymptotic energy is -1.816 eV.}
    }
    \label{fig:H3pH_compressed}
\end{figure}

The more accurate description of the higher energy regime, energies larger than 2 eV,
can be seen in Figure~\ref{fig:H3pH_compressed} where H approaches to a compressed H$_3^+$.
The differences between PES1 and PES-NN are more pronounced when bonds are compressed than stretched. This is

\section{Reaction dynamics}
  
\subsection{Quasi-classical  trajectory and surface hoping  calculations}

The quasi-classical trajectory calculations are performed with the  MDwQT
code \cite{Sanz-Sanz-etal:15,Zanchet-etal:16,Ocana-etal:17,Roncero-etal:18}.
When considering several coupled adiabatic electronic states, the fewest switches approach of Tully \cite{Tully:90}
is used as described in \cite{Sanz-Sanz-etal:15}.
Initial conditions are sampled with the usual
 Monte Carlo method \cite{Karplus-etal:65}.
The initial conditions for the vibrational modes
of H$_2$ and H$_2^+$ are quantized using the adiabatic switching method  \cite{Grozdanov-Solovev:82,Qu-Bowman:16,Nagy-Lendvay:17},
yielding vibrational energies within 0.3 meV with respect to the exact vibrational levels of each diatomic fragment.
The rotational states of  H$_2$ and H$_2^+$ are set to zero in these studies, and the initial distance
between the two center-of-mass
is set to 105 bohr. The initial impact parameter, $b$, is sampled between 0 and B, according to a quadratic distribution on $b$, where
$B$ is determined for each energy according to a capture model \cite{Levine-Bernstein:87} as 
$    B=\left( \alpha / 2E\right)^{1/4}$,
for a charge-induced-dipole interaction,
{
described by  $-\alpha/2R^4$, where $\alpha$ is a
constant,  with dimensions
$\lbrack E L^4\rbrack$, which is proportional to the average
polarizability of H$_2$ , $\beta$, at its equilibrium configuration, 
as $\alpha= \beta e/4\pi\epsilon_0$.
}
All trajectories are stopped when any internuclear distance becomes longer than 125 bohr, where they
are analysed.

The reactive cross section
for each  collision energy, $E$,  is  calculated  as \cite{Karplus-etal:65}
\begin{eqnarray}\label{qctXsection}
\sigma_{vv'}(E)= \pi b_{max}^2 P_r(E)
 \quad {\rm with }\quad
  P_r(E)= {N_r\over N_{tot}},
\end{eqnarray}
where
$N_r$ is the number of trajectories leading to products
and $N_{tot}$ is the total number of trajectories with initial impact parameter lower
than $b_{max}$, the maximum impact parameter for which reaction takes place at energy $E$.
Here we have considered H$_2$($v$=0), while H$_2^+$ vibrational level $v'$ varies between 0 and 6.
For each ($v,v'$) couple  and each energy a set of 10$^5$ trajectories are run, with
energy error lower than 0.01 meV.

\begin{figure}[t]
\begin{center}
 {
 \includegraphics[scale=0.4]{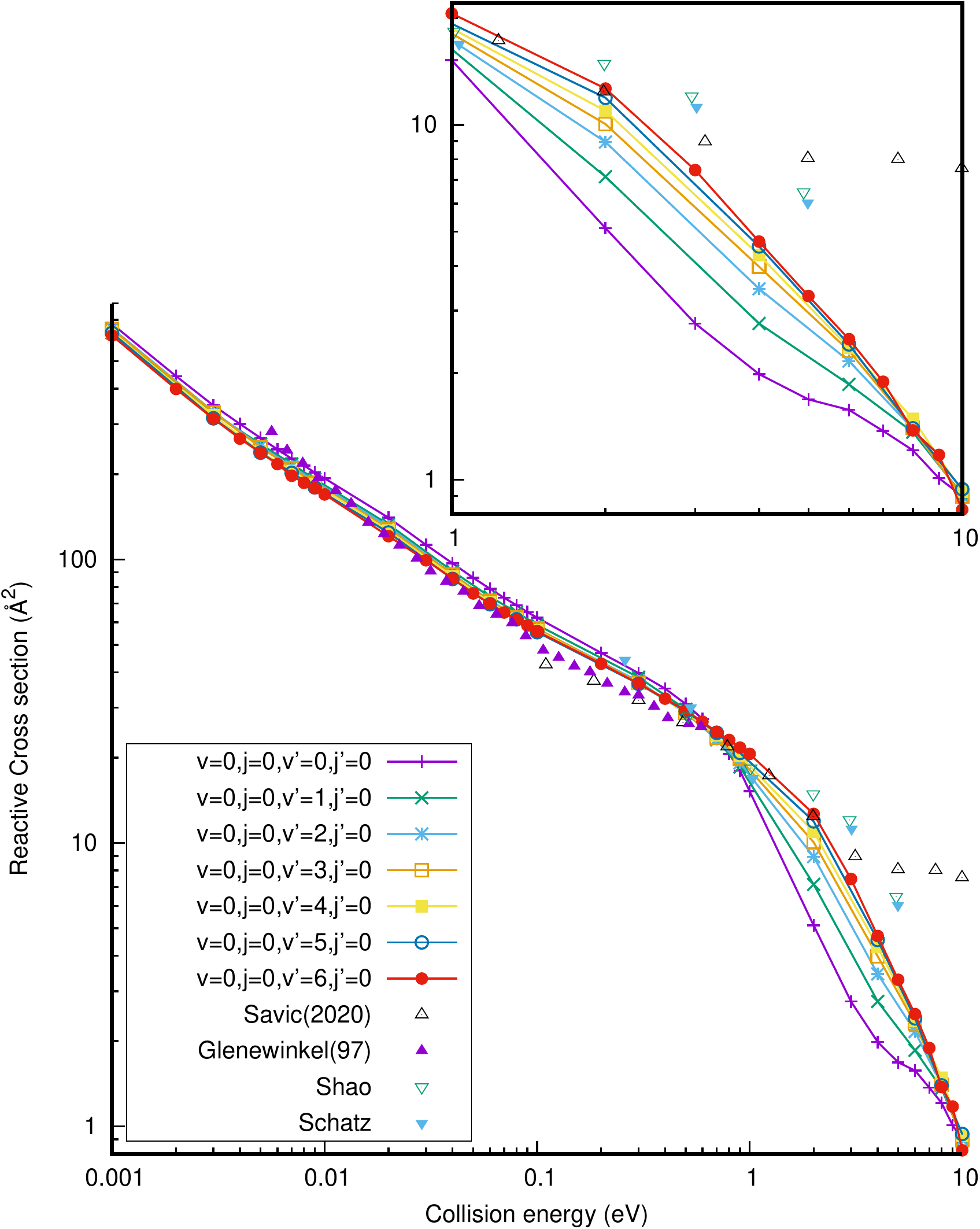}
 \caption{\label{H2yH2p-vibrational}{\it  H$_2$($v$=0,$j$=0) + H$_2^+$($v'$,$j'$=0) $\rightarrow$ H$_3^+$ + H
     reactive cross sections
     obtained with QCT calculations for different vibrational states $v'$ of H$_2^+$. The experimental
     results are those of Savic {\it et al.} \cite{Savic-etal:20}, from Glenewinkel-Meyer and D. Gerlich \cite{Glenewinkel-Meyer-Gerlich:97} and
     Shao and Ng \cite{Shao-Ng:86}, and the theoretical results of  Eaker and  Schatz \cite{Eaker-Schatz:85}.
}}
}
\end{center}
\end{figure}

{ 
  The final energy distribution of H$_3^+$ products is also analyzed. We simply evaluate classical
  energies, without trying to consider the permutation symmetry of identical fermions (for H$_4^+$)
  or bosons (for D$_4^+$). This is done in three steps. First, the kinetic energy of H$_3^+$ and H products
  are calculated and substracted. Second, the rotational angular momentum of H$_3^+$ products is evaluated, and 
  its rotational energy. By setting the origin of energy at the bottom of the H$_3^+$ well, the remaining energy
  corresponds to vibrational energy. Here, we do not attempt to assign the internal vibrational modes, which deserves
  further development and is led for a future work.

}  
\subsection{{\rm H}$_2$ {\rm +} {\rm H}$_2^+$($v'$) collisions in PES1}

The new experimental results for the title reaction of Savic {\it et al.} \cite{Savic-etal:20} differ
from previous ones, both experimental and theoretical, above 2 eV. The difference with previous
experimental data, which are scattered above 1-2 eV, may be due to different conditions in the generation
of the reactants. In low temperature plasma the vibrational temperature of H$_2$ is of the order of 2500 K,
so that the population of H$_2$($v$=1) is expected to be lower
than 10\% \cite{Mendez-etal:06,Jimenez-Redondo-etal:11,Tanarro-Herrero:11}. On the contrary, H$_2^+$ is formed
by electronic impact or photoionization, which  may yield to different
vibrational and rotational excitations. Vibrationally excited H$_2^+$
can partially  thermalize, yielding to different initial conditions in different experimental setups.
As an example, recent theoretical calculations \cite{Sanz-Sanz-etal:21,Roncero-etal:22}
on the H + H$_2^+$  charge transfer reaction, and some isotopic
variants, have found that the reaction cross section highly depends on the initial vibrational 
state of the diatomic ion. Following this idea, in this work we have performed QCT calculations
of the cross section of the reaction for several initial vibrational states
of H$_2^+$ reactant using the global PES1 of Refs. \cite{Sanz-Sanz-etal:13,Sanz-Sanz-etal:15}, which are shown
in Figure~\ref{H2yH2p-vibrational}.

For exothermic reactions without a barrier, the long-range interaction between the reactants dominates the reaction dynamics.
In the case of charge-induced dipole  long-range interactions
the cross section  for exothermic reactions takes the form \cite{Levine-Bernstein:87}
\begin{eqnarray}\label{Langevin-E-dependence}
\sigma(E)= \pi \, (\alpha/E)^{1/2}.
\nonumber\\
\end{eqnarray}
This is approximately the behaviour of the cross sections for energies below 1 eV for
every initial vibrational state $v'$. Therefore,
we can conclude that in this energy interval reaction dynamics is dominated by
long range interactions, independently of the initial vibrational state of H$_2^+$($v'$).

\begin{figure}[t]
\begin{center}
  {
    \includegraphics[scale=0.4]{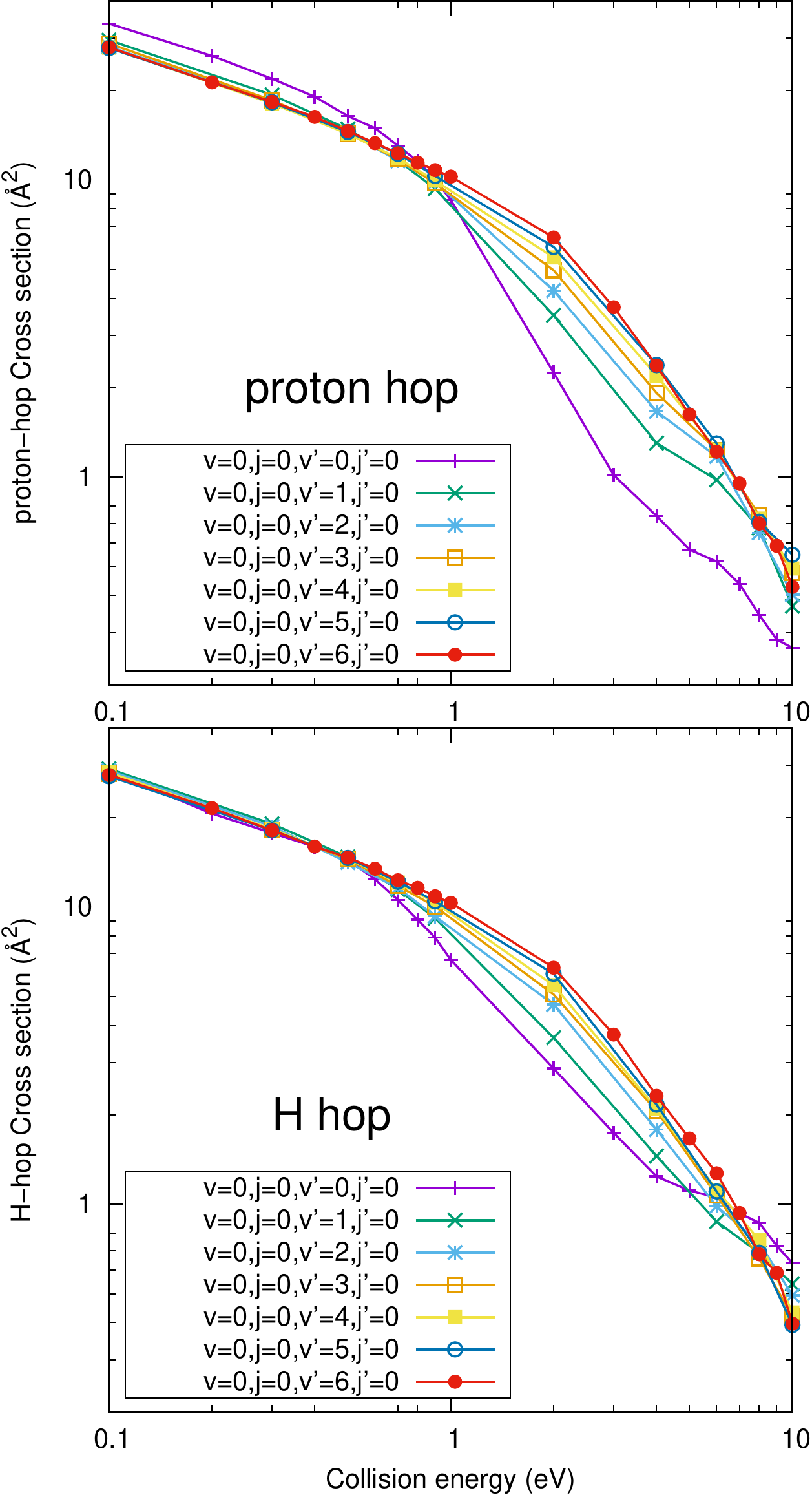}
    \includegraphics[scale=0.4]{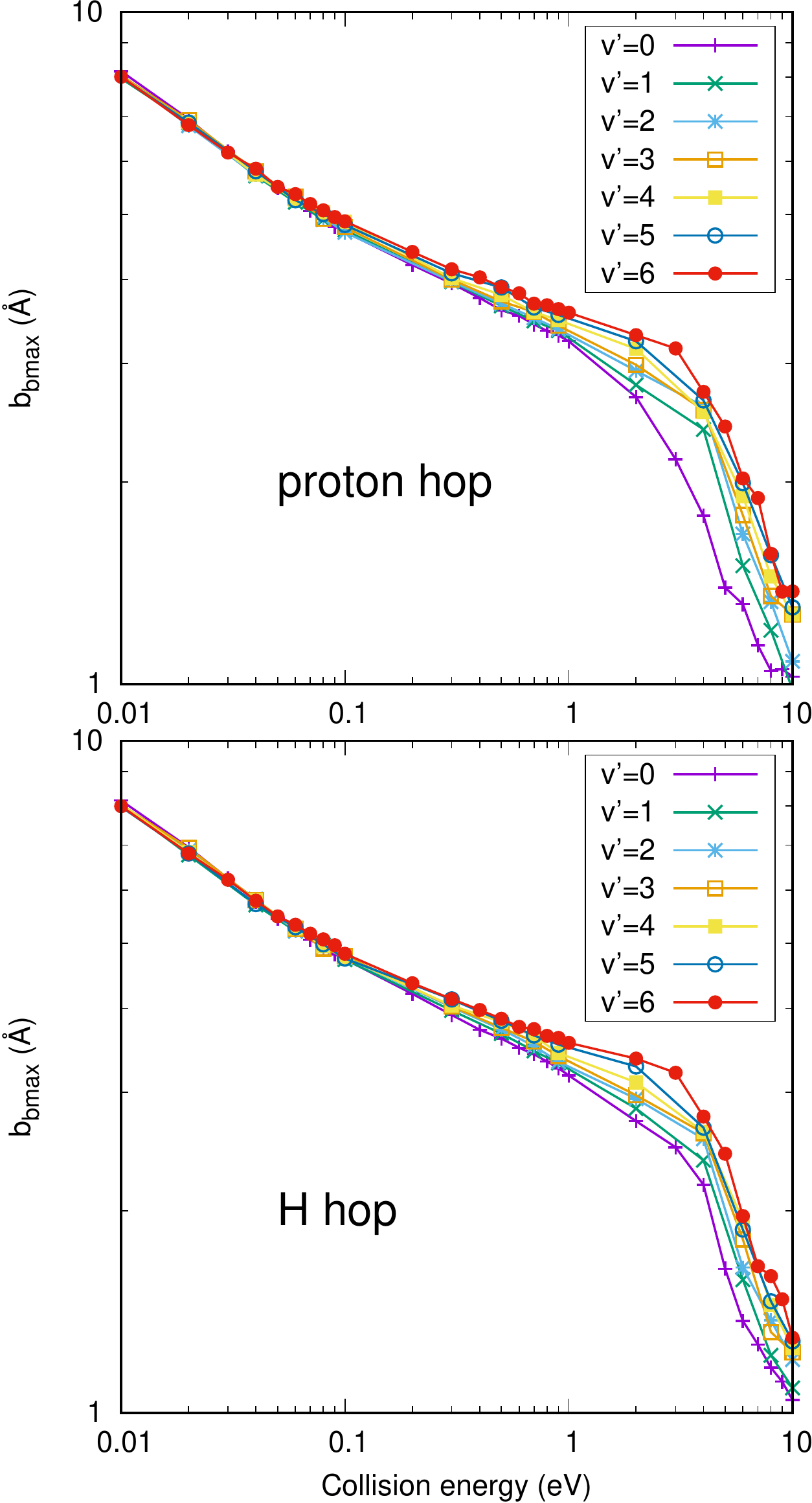}
    \caption{\label{H2yH2p-protonYh-hop}{\it  H-hop and proton-hop cross sections (left panels)
        and maximum impact parameter, $b_{max}$, (right panels)
     for the H$_2$($v$=0,$j$=0) + H$_2^+$($v',j'$=0) reaction
     obtained with QCT calculations for different vibrational states $v'$ of H$_2^+$.
     
}}
}
\end{center}
\end{figure}

However, above 1 eV the reactive cross sections present important differences among the  $v'$ considered,
showing that the vibrational
excitation has a strong impact on the reactivity. In general,  the reactive cross section
increases with increasing $v'$,
which could  be simply understood assuming that H$_2^+$ can break more easily. In the left panels of
Figure~\ref{H2yH2p-protonYh-hop}
the two main mechanisms to form H$_3^+$ are separated as H-hop and proton-hop, corresponding to the fragmentation
of H$_2$ or H$_2^+$ reactants, respectively. For collision energies below 1 eV, the cross section
for the proton-hop mechanism
is slightly larger. However, above 3 eV the H-hop cross section is larger for $v'$=0, and the two mechanisms
tend to a rather similar value as $v'$ increases. This means that the vibrational excitation
of H$_2^+$ does not produce significant increase of
the proton-hop mechanism. This can be explained looking at the right panels of Figure~\ref{H2yH2p-protonYh-hop},
where the maximum impact parameter, $b_{max}$, is plotted for each proccess and inital vibrational
state. Above 1 eV,  $b_{max}$ increases
from $v'$= 0 to $v'$=6, in nearly an identical quantity for the two mechanisms. We conclude
that the increase of the cross section when varying $v'$ is due to the growing of the H$_2^+$ subunit,
whose right turning point increases from 1.25 to 2.12 \AA, for $v'$=0 and 6 respectively. Therefore, at energies
above 1 eV, when long-range interactions are not able to produce important deviations among the reactants,
the size of the two reactants approximately determines the value of the maximum impact parameter.
At these higher energies a more complex reaction
mechanism occurs, in which H$_2$ may nearly insert in the H$_2^+$ bond, specially at high $v'$ excitations.

\subsection{{\rm D}$_2$ {\rm +} {\rm D}$_2^+$ collisions in PES1}
\begin{figure}[t]
\begin{center}
 {
 \includegraphics[scale=0.4]{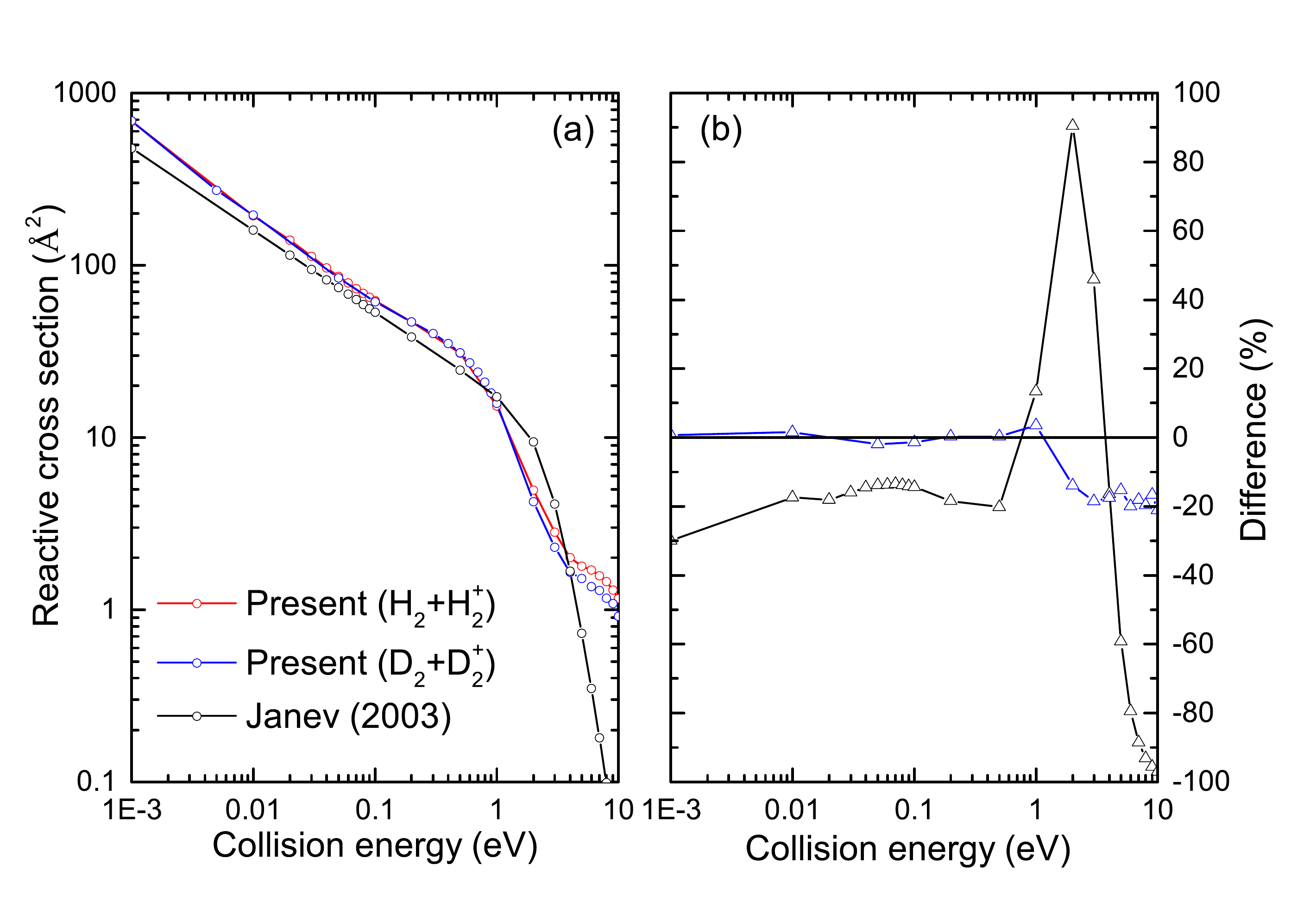}
 \caption{\label{D2yD2p-v0-vp0}{\it
(a) D$_2$($v$=0,$j$=0) + D$_2^+$($v'$=0, $j'$=0) reactive cross sections (blue line with open circles)
     compared with those for H$_2$($v$=0,$j$=0) + H$_2^+$($v'$=0,$j'$=0) (red line with open circles)
     obtained by QCT calculations. The cross section from Janev et. al. \cite{Janev2003} {  for  H$_2$ + H$_2^+$} reaction}
     is also displayed for comparison (black line with open circles).
     (b) The difference of each cross section relative
     to the present H$_2$ + H$_2^+$ cross section (open triangles on the colored line).
}}

\end{center}
\end{figure}

The reactive cross section for the D$_2$($v,j$) + D$_2^+$($v',j'$) collisions is
shown in Figure~\ref{D2yD2p-v0-vp0} for $v$ = $j$ = 0 and $v'$ = $j'$ = 0. For energies below 1 eV, the results
for D$_2$ + D$_2^+$ closely match those for H$_2$ + H$_2^+$. This can be explained
in terms of the Langevin model, in which the cross section of Eq.~(\ref{Langevin-E-dependence}) does
not depend on the mass of the reactants. This explains why the  reaction cross sections 
  for  D$_4^+$ and H$_4^+$ are nearly the same.

However, it should be noted that for reactions involving partially deuterated species, the reaction
cross section presents larger differences, as those already reported in the theoretical study
of Ref.~\cite{Sanz-Sanz-etal:15}. The shift of the center-of-mass with respect to the geometric
center of the two diatomic reagents introduces important differences, specially related to 
the effect of the rotation. In particular, in the homonuclear
neutral H$_2$ or D$_2$ case for $j$=0, only the charge-electric dipole term affects  the long-range
interaction, determining the reactive cross section below 1 eV. In this case, the charge-electric quadrupole
term vanishes when integrating over the angular coordinate for the isotropic $j$=0 case (but not for $j>$ 0).

\subsection{Non-adiabatic effects and fit accuracy}

The increasing of the H$_2^+(v')$ vibrational excitation yields to an increase
in the reactive cross section. However, this increasing, even for $v'$=6 does not match the
new experimental data by Savic {\it et al.} \cite{Savic-etal:20}. Since the cross section as a function
of the $v'$ excitation seems to converge to the value of $v'$=6, here after we shall focus on
the two limiting cases, $v'$= 0 and 6.

\begin{figure}[t]
\begin{center}
 {
 \includegraphics[scale=0.4]{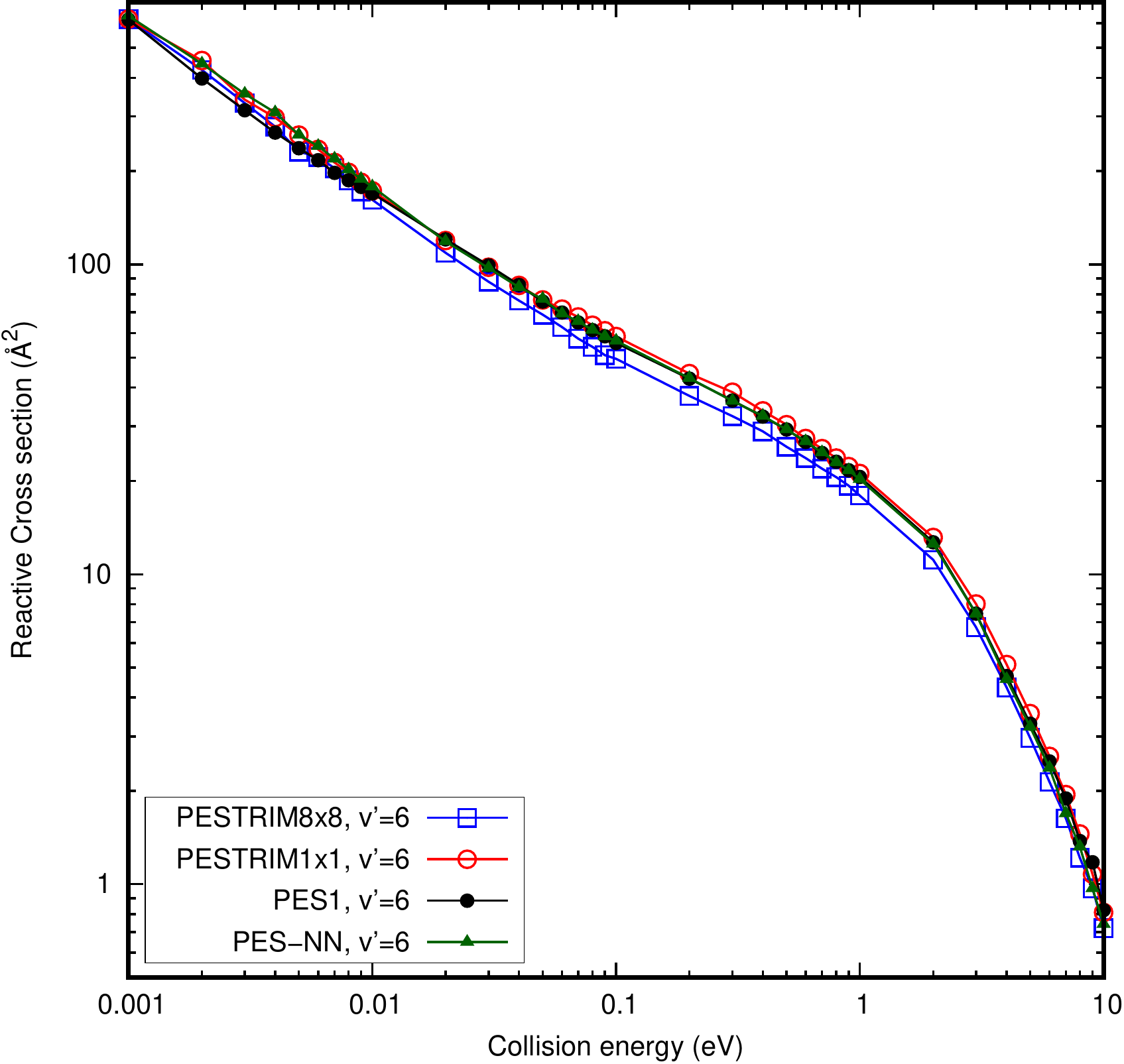}
 \caption{\label{non-adiabatic-sigma}{\it  QCT H$_2$($v$=0,$j$=0) + H$_2^+$($v'$=6,$j'$=0) $\rightarrow$ H$_3^+$ + H
     reactive cross sections obtained
     with the PESTRIM8$\times$8 using surface hoping method (blue),  the PESTRIM1$\times$1 (red), the PES1 (black)
     and the new PES-NN (green) potential energy surfaces.
}}
}
\end{center}
\end{figure}

In order to investigate
the role of electronic transitions among the lower electronic states, in this work we use
the PESTRIM8$\times$8, comparing it to the results on  the PESTRIM1$\times$1 model.
The dynamical results obtained for
these two potentials are shown in Figure~\ref{non-adiabatic-sigma}, and compared with the data
obtained with PES1. All the results present a similar behaviour, with small differences
in the logarithmic scale used in the figure. All the results converge to the same value
at the lowest collision energy considered here, 1 meV, since all the PESs used in this work have
similar long range interactions.

At intermediate energies, between 0.01 and 2 eV, the PESTRIM8$\times$8 and  PESTRIM1$\times$1 cross sections
are slightly different, showing a small effect of non-adiabatic transitions in the dynamics. Curiously,
these non-adiabatic effects seem to produce a decrease on the reactive cross section, in contrast to
what is needed  to match the values at higher energies of the experimental results of Savic {\it et al.} \cite{Savic-etal:20}.

In  figure\ref{non-adiabatic-sigma} we also compare with the cross section obtained with the adiabatic PES1, which nearly matches
the results obtained with  the adiabatic  PESTRIM1$\times$1 up to 3 eV. Above 3 eV, the results obtained with  PES1 are higher than
those for PESTRIM1$\times$1, showing that the four body term may be important. However, for the new PES-NN, of higher
accuracy than PES1, the reaction cross section is nearly identical to that of PESTRIM1$\times$1 for collision energies
below 1 eV and very close to those obtained with PESTRIM8$\times$8 for energies above 4 eV. From this comparison
we may conclude that the better accuracy of the four body term of the new PES-NN does not improve significantly
the differences between the simulated and measured \cite{Savic-etal:20} reactive cross-sections.

In order to analyze whether there are other excited states not well described by the TRIM approximation, we have performed
{\it ab initio} calculations of the four lower electronic states, considering a larger electronic basis set, with
extra orbitals added to describe the 2s and 2p electronic states of atomic hydrogen. With this larger basis set,
we have found that the energies obtained (extrapolated to the complete basis set) differ only a few tenths of meV
with the previous {\it ab initio} calculations, and that no higher electronic state appears below 10 eV.

\subsection{Double fragmentation and reaction mechanism}

\begin{figure}[t]
\begin{center}
 {
 \includegraphics[scale=0.8]{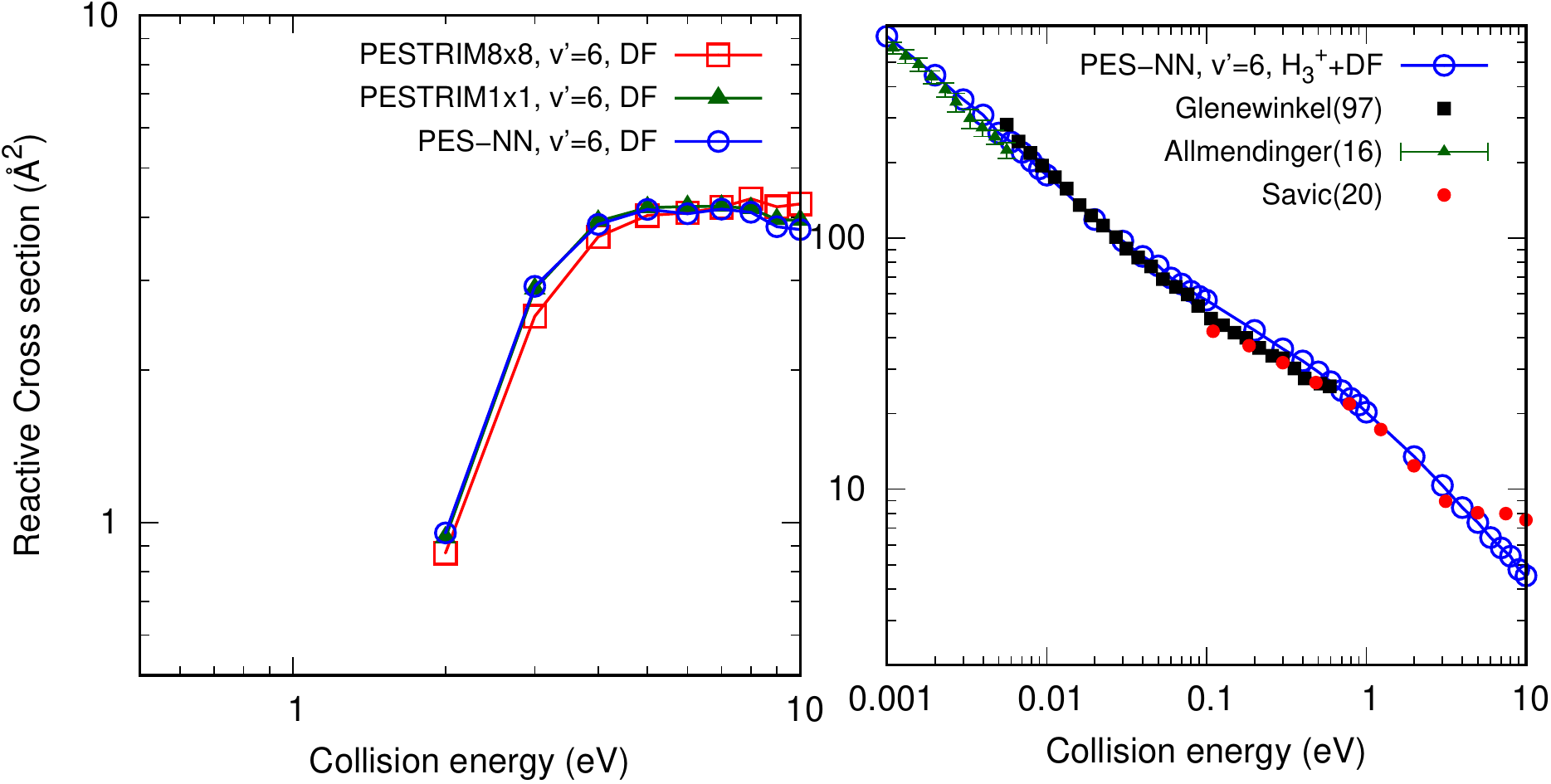}
 \caption{\label{Double-fragmentation}{\it  QCT H$_2($v$=0,$j$=0)$ + H$_2^+($v'$=6,$j'$=0)$ $\rightarrow$ H$_2$ + H$^+$ + H (DF channel)
     reactive cross sections (left panel) and total cross section (formation of H$_3^+$ and DF channel, in right panel) obtained
     with the PESTRIM8$\times$8 using surface hoping method (blue),  the PESTRIM1$\times$1 (red)
     and the new PES-NN ( green) potential energy surfaces. The experimental
     results are those of Refs. \cite{Glenewinkel-Meyer-Gerlich:97}, \cite{Allmendinger-etal:16b} and \cite{Savic-etal:20} and .
}}
}
\end{center}
\end{figure}

 For H$_2^+(v'=6)$
 the double fragmentation (DF) channel opens at $\approx$ 2 eV, as it is shown in the left panel of Figure~\ref{Double-fragmentation}.
 The opening of this channel occurs approximately for values where the extra energy of $v'$=6, 1.67 eV,
 is added to the D$_0$ of H$_2^+$, 2.65 eV

If we add the DF channel contribution to the production of H$_3^+$, as shown in the
right panel of Figure~\ref{Double-fragmentation}
the cross section increases considerably, reaching a very good agreement
with the new experimental data of Savic {\it et al.} \cite{Savic-etal:20}
above 3 eV.

\begin{figure}[t]
 {
 \includegraphics[scale=0.4]{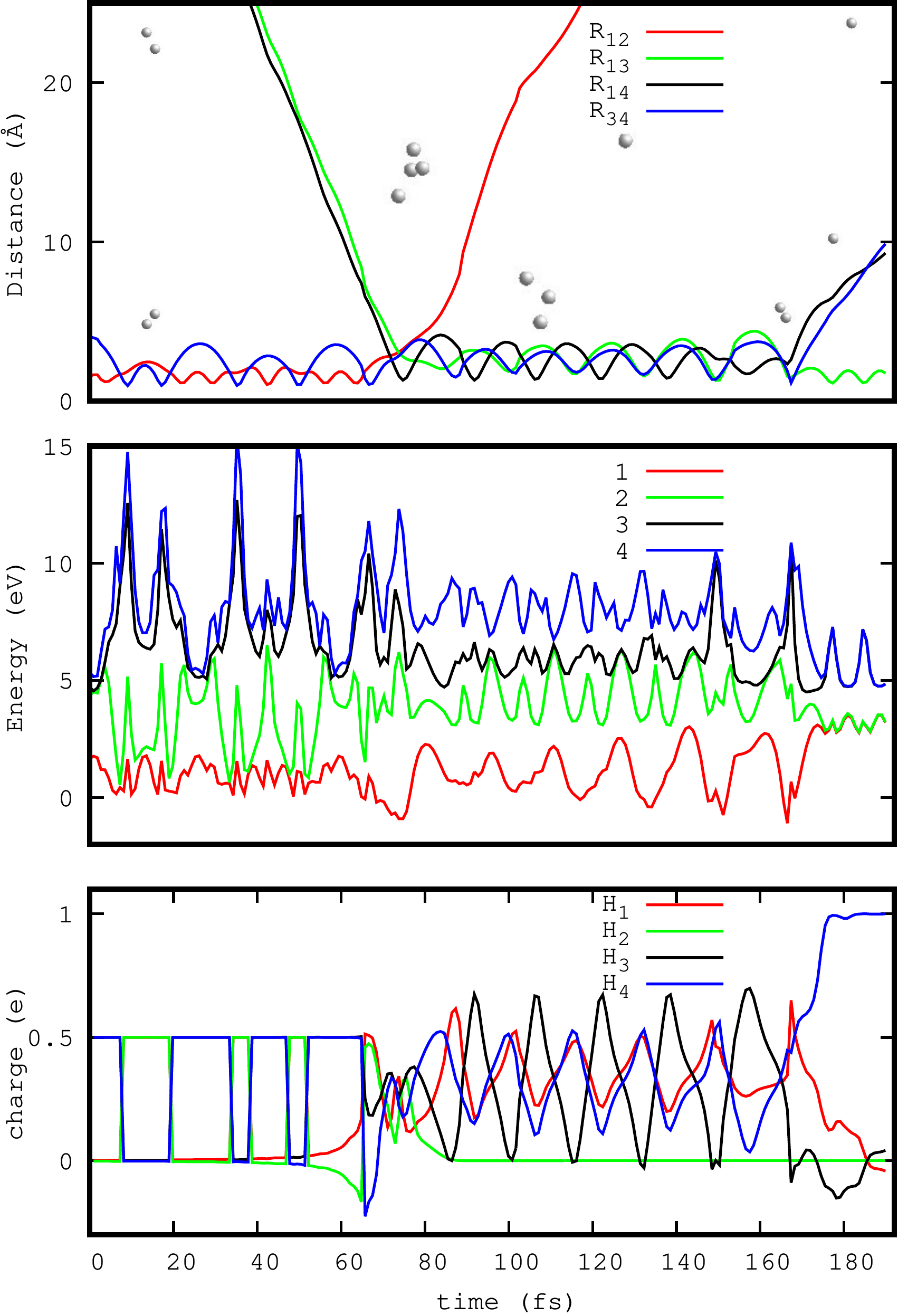}
 \includegraphics[scale=0.4]{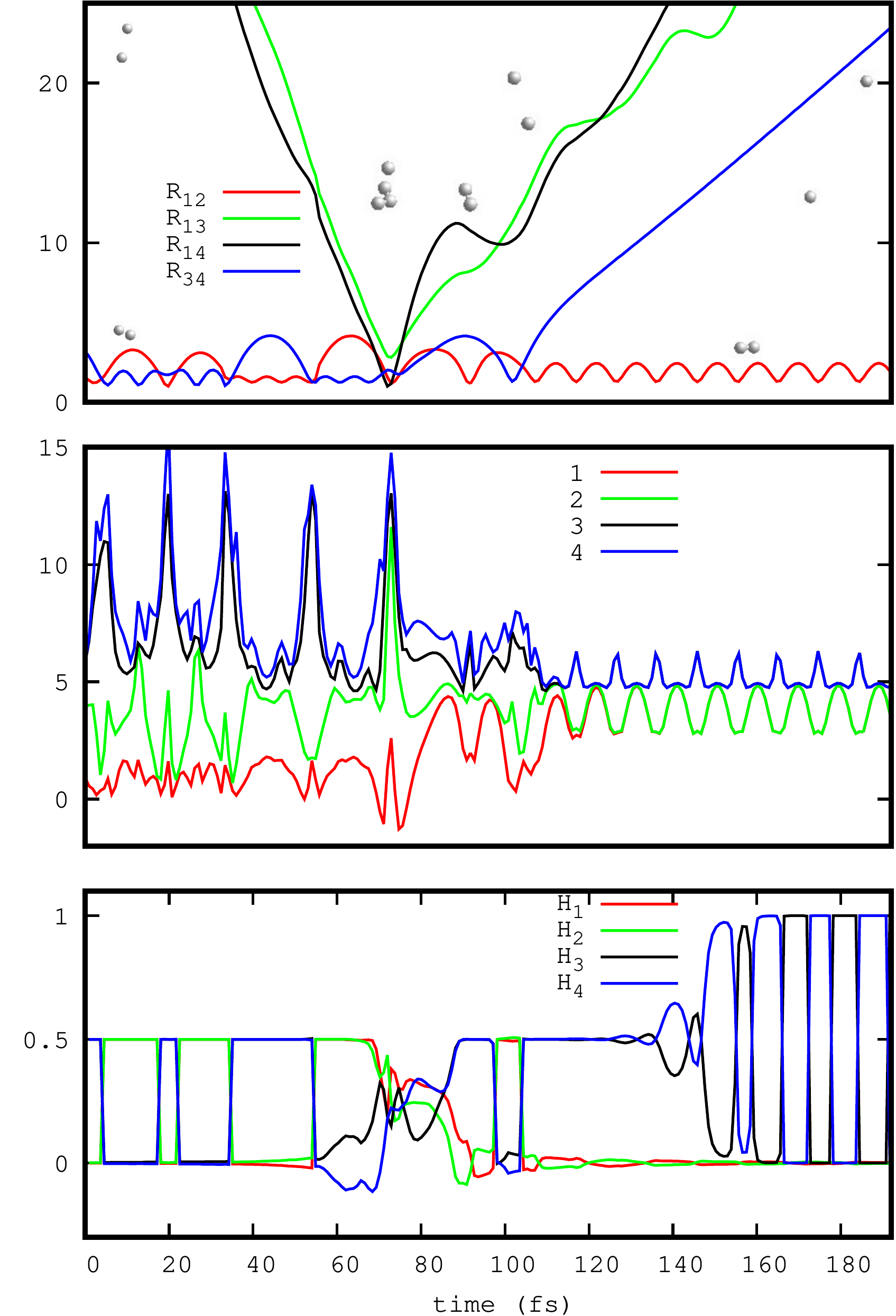}
 \caption{\label{trajectory-Double-fragmentation}{\it Two typical trajectories leading to double fragmentation (DF),
     as a function of the collision time in fs.
     Lower panels show the charge on each atom (Mulliken population) for the ground electronic state. Middle panels
     shows the energies of the four lower electronic states. Upper panels show some characteristic internuclear distances
     needed to characterize the trajectory. { $R_{ij}$ refers to the distance between atom H$_i$ and H$_j$}
}}
}
\end{figure}

The main mechanism giving rise to the DF channel consists of three steps,
as it is shown in the left panels of Figure~\ref{trajectory-Double-fragmentation}.
First, a highly vibrationally excited H$_4^+$ intermediate is
formed by insertion of H$_2$ in the elongated H$_2^+$, which lives short time.
Second, a first H atom is ejected (in the Figure is atom 2 with charge 0),
forming a very excited (H$_3^+$)$^*$, which lives from 80 to 160 fs, approximately. In the third step,
one of the atoms of H$_3^+$ (atom 4 in the Figure) dissociates, carrying the positive charge, thus
leading to neutral H$_2$.
Since this third step occurs much later, it could explain that experimentally the metastable  (H$_3^+$)$^*$
would be detected
together with more stable H$_3^+$ products.  It is also important to notice
that the energy transfer among particles of identical mass is possibly overestimated
in a classical treatment, as the one used in this work. If this is the case, it would
 support  the inclussion of the DF cross section as a part of the total reaction
cross section. Quantum calculations are needed to solve this problem, but they are
rather challengig at the high energy considered. 

The right panels of  Figure~\ref{trajectory-Double-fragmentation} show another DF mechanism: in this
case H$_2$ and H$_2^+$ are produced at 70-80 fs.
H$_2^+$ is vibrationally very excited  and dissociates
later, at $\approx$ 100-110 fs. In this case, there are two
degenerate electronic states, each one corresponding to the charge in one of the ejected atoms, at long
distances from the H$_2$ fragment. In the ground electronic state, shown in the lower panels,
the charge is exchanged between the two identical atoms, 
because of this degeneracy, and shows the nature of the surface hopping occurring in the products channel
when including
several electronic states (PESTRIM8$\times$8). The electronic transition occurs among degenerate electronic
states describing H$_2$ + H + H$^+$ and H$_2$ + H$^+$ + H products, what explains the small effect of including
electronic transitions in the reaction dynamics.

In addition, the charge transfer in the entrance channel  occurs
between H$_2$ and H$_2^+$, when the two reactants have the same internuclear distance
(see bottom panels of Figure~\ref{trajectory-Double-fragmentation}). In this
situation there is a degeneracy between the
two lower adiabatic states, as discussed in detail in Ref.~\cite{Sanz-Sanz-etal:15}.

\begin{figure}[t]
 {
 \includegraphics[scale=0.7]{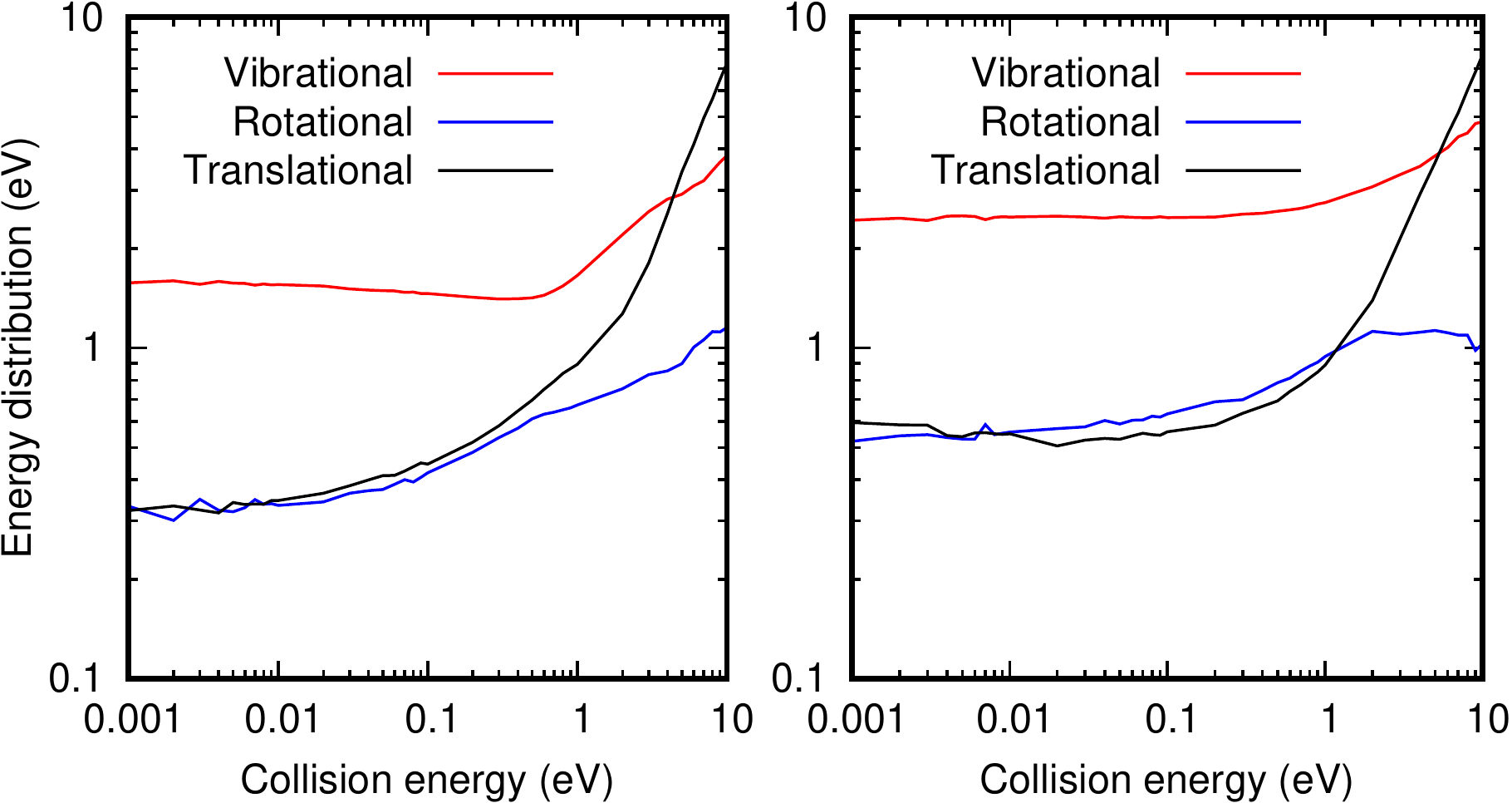}
 \caption{\label{EnergyDistribution}{\it Vibrational, rotational and translational energy
     distributions of the H$_3^+$ + H products, as a function of the collision energy for the H$_2 (v=0)$
     + H$_2^+ (v')$ collisions, for $v'$=0 (left panel) and $v'$=6 (right panel).
     { The origin of energy is in the bottom of the well of the H$_3^+$ products}.
     The initial vibrational energies of the reactants is 0.413 and 1.673
     eV for the $v'$= 0 and 6, respectively, { with respect to the minimum of each fragment}.
    { The potential energy difference between reactants and products
     is 1.816 eV, and when ZPE are accounted for, the exoergicity of the present PES becomes 1.688 eV}.
}}
}
\end{figure}

The energy distributions of the H$_3^+$ + H products are shown in Figure~\ref{EnergyDistribution}, and it is nearly
quantitatively the same for all the PESs.
The energy difference between the two vibrational states of  H$_2^+$($v'$= 0 and 6) is approximately
1.4 eV, close to the exoergicity. For low collision energies, 0.1-1 meV, the initial vibrational
energy of  H$_2^+$( $v'$= 6) is  1.40 eV higher than that of H$_2^+$ ($v'$= 0), and the vibrational energy
of the corresponding H$_3^+$ products is also higher, but only by $\approx$ 0.9 eV.  
Therefore the remaining 0.4 eV are nearly equally distributed between the rotational and translational energies
of the products. Rotational energy increases with collision energy,
except for the $v'$= 6  above 1 eV, where rotational excitation reaches a plateau and seems
to start decreasing. The translational energy always increases for $v'$= 0, while
for $v'$= 6 it slightly decreases below 0.1 eV, and then increases again.
The vibrational energy of products shows two different behaviors:
 below $\approx$ 0.7 eV, the vibrational energy
 slightly decreases ($v'$=0)  or remains constant ($v'$=6); above this energy the vibrational
 energy increases sharply.

 Such behavior allows to assign two different reaction mechanisms. Below $\approx$ 0.7 eV,
 the impact parameter (in Figure~\ref{H2yH2p-protonYh-hop}) is large, supporting
 a stripping mechanism, in which the long range interactions attract the reactants to each
 other, originating orbits enhancing the relative angular momentum between the two reactants.
 Above $\approx$ 0.7 eV, however, the impact parameter is rather small, and the reaction
 occurs by an insertion mechanism, in which the H$_3^+$ is greatly excited vibrationaly, specially
 as initial vibrational and translational energy of the reactants increases.

 The
 {   dissociation energy  of H$_3^+$ is 4.34 eV, very close to the value reported by \cite{Mizus-etal:17} of 4.35 eV,
 and the average vibrational energy distribution
 of H$_3^+$ reaches values in the  4-5 eV interval, $i.e.$ values above the dissociation energy explaining
 why  H$_3^+$  dissociates, leading to the DF channel.
}

\section{Plasma modelling}

 To model the hydrogen plasma, we shall consider a gas-discharge vessel of cylindrical symmetry, so that
  only $z$, parallel to the central axis, and $R$, the distance from the center to the walls, will be considered,
  with the cylinder of infinite length in this case. The gas is initially in the form of neutral H$_2$, and after the discharge ignition new
  species are formed, H, H$^+$, H$_2^+$,  H$_3^+$ and electrons. H$^-$ is neglected under the conditions considered
  here. To model the abundance of these species in the
  stationary condition we use a model similar to those
  already described previously \cite{Lishev,Koleva-etal:03}, which is outlined in the appendix \ref{ap:plasma},
  including all the processes listed in Table~\ref{tab:reactions}, in which the vibrations
  of molecular species, H$_2$, H$_2^+$ and H$_3^+$ are not considered.

  The plasma model is done for
  pure hydrogen and pure deuterium gases. For modelling deuterium plasma, the cross sections for electron collisions
  and radiative transitions of D species are used by those of H species.
  
  The  cross section for  H$_2$ + H$_2^+$ and  D$_2$ + D$_2^+$ reactive collisions calculated in this work
  are included in the models presented below. The plasma modellings are also done using the cross section
  by Janev {\it et al.} (2003) \cite{Janev2003}, which are compared with the present calculations in Figure~\ref{D2yD2p-v0-vp0}.

  Here, the electron temperature $T_e =  5.55$~eV and electron density $n_e = 2.07 \times 10^{12}$ cm$^{-3}$ are used,
  which were measured  by a Langmuir probe for D plasma in a long cylindrical vessel \cite{Chai}. The molecular temperature
  is set to $T_m$= $0.026$~eV (300 K) and the atomic temperature is   $T_a$ =  $0.052$~eV (600 K).
  At these realistic temperatures, the relevant energies are below  $E_{col}$= 1 eV, so that the H$_2^+$ vibrational
  excitation has no significant effect.

\begin{figure}[t]
    \centering
    \includegraphics[scale=0.5, trim={0.5cm 0.5cm 0.1cm 0.1cm}]{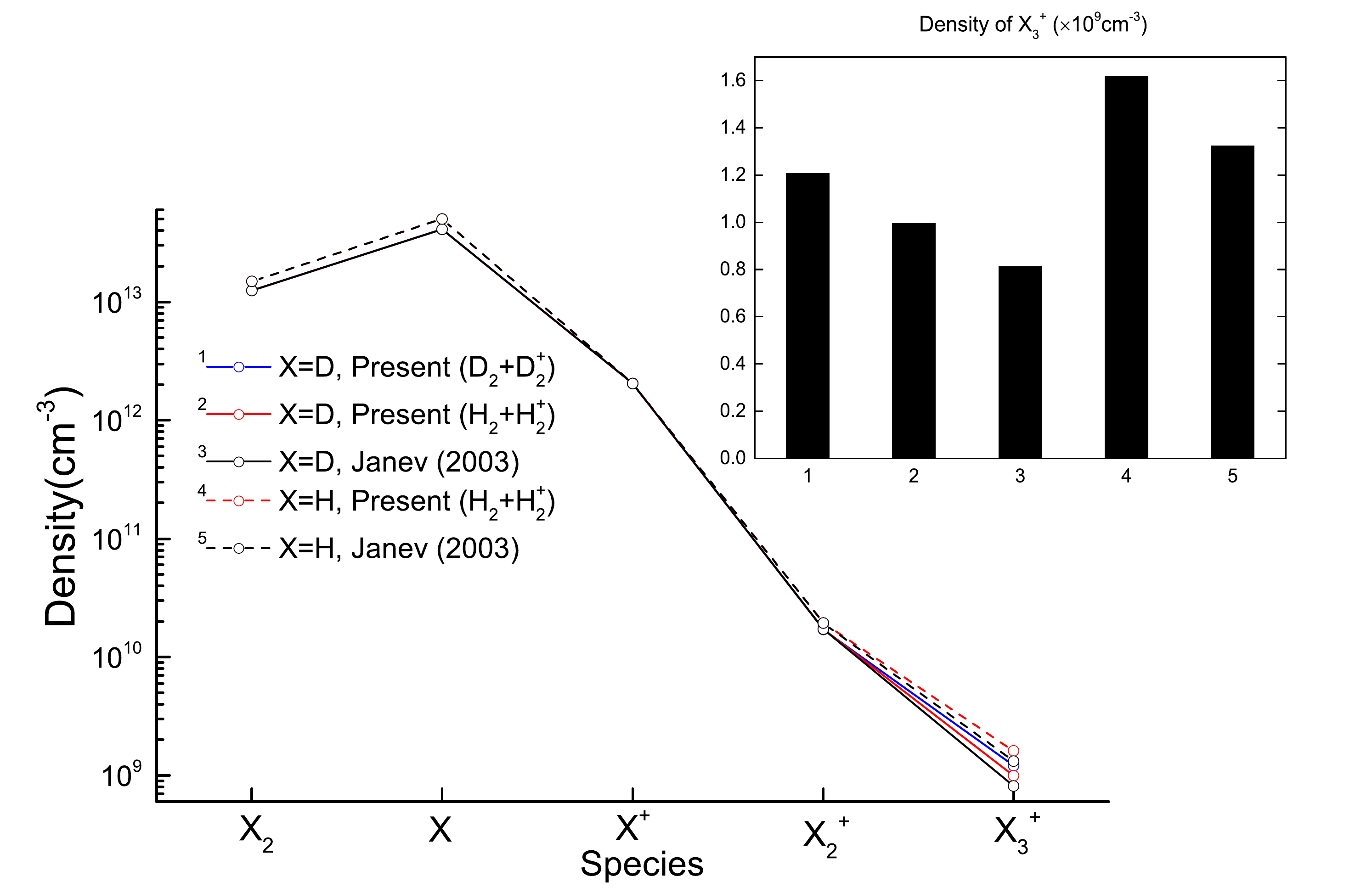}
    \caption{The population densities of D and H
      species by modeling using the different cross sections shown in 
      Figure \ \ref{D2yD2p-v0-vp0}. In the inset, the labels of x-axis correspond
    to the cross sections used given by the upper number in the legends of the main graph.} 
    \label{fig:D3+density}
\end{figure}

Figure\ \ref{fig:D3+density} shows the resulting population densities of D and H species depending on
the cross section used. 
We can note, the  D$_3^+$ and H$_3^+$ population densities vary significantly with the cross section used,
while other species are nearly unchanged, 
as shown in Figure\ \ref{fig:D3+density}. The  D$_3^+$ and H$_3^+$ population density changes 
are also summarized in Table~\ref{tab:D3+density}.
It is worth noting  that the main depopulating mechanism for H$_3^+$ density is the diffusion process
(given in  the last term of Eq.\ \ref{balance1}), while  the electron impact processes (H$_3^{+}+e$ in Table\ \ref{tab:reactions})
only contribute to the depopulation in a small fraction.

The present cross section for D$_2$ + D$_2^+$ is larger than that 
for H$_2$ + H$_2^+$ by $\sim 2$\% at low collision energy, below $\sim 0.026$~eV. On the contrary
for collision energies above 1 eV, the cross section for deuterium is $\approx$ 20$~\%$
lower than that of pure hydrogen reaction,
as shown in Figure~\ref{D2yD2p-v0-vp0}.b. For H$_2$ + H$_2^+$,  the cross section
by Janev et. al \cite{Janev2003} is smaller than the present ones by
$20\sim30$~\% below 1 eV, but the difference
is enlarged up to 100~\% (black triangle) at the collision energy over $\sim 1.0$~eV
as shown in Figure~\ref{D2yD2p-v0-vp0}.b.

The reaction rate coefficient calculated at the molecular temperature, $T_m=0.026$~eV,
for  pure deuterium  is $\approx$ 20$~\%$ larger than for pure hydrogen, and is also larger
than those obtained from Janev {\it et al.} \cite{Janev2003} by $\sim40$~\%, as listed in Table~\ref{tab:D3+density}.
These differences have a rather linear impact on the resulting   D$_3^+$ and H$_3^+$ population densities,
whose difference varies proportionally to the difference
between the rate coefficients  listed in  Table\ \ref{tab:D3+density}. 

As a result, the use of the present results for D$_2$ + D$_2^+$ and H$_2$ + H$_2^+$ leads
to significantly different D$_3^+$ and H$_3^+$ population densities  compared
with the widely used cross section for H$_2$ + H$_2^+$ \cite{Janev2003} in this plasma modeling.
It should also be noted that the use of the cross section for H$_2$ + H$_2^+$ instead of
that for  D$_2$ + D$_2^+$ in the modeling of D plasma can give rise to unreliable
population density of D$_3^+$, even though the difference between the two cross sections is small,
due to the rate coefficient sensitivity to the cross section at the low collision energy. 


  When T$_e$ and n$_e$ are reduced to 3 eV and 1.5 $\times$ 10$^{10}$  (cm$^{-3}$),
  respectively, and the molecular pressure is increased by about 10 times,   
  the amount of X$_3^+$ (X=D, H) becomes dominant over those of X$_{1,2}^+$ ions.
  These conditions are similar to those reported by
  Tanarro and Herrero  \cite{Tanarro-Herrero:11}, who also
  find a significant increase of the H$_3^+$ population.
  This relative increase of the  X$_3^+$ density is mostly attributed to that
  the rate coefficient of X$_2$ + e collision populating X$_2^+$ becomes smaller than
  the rate coefficient of X$_2$ + X$_2^+$ collision depopulating X$_2^+$.
 
 However, the density of X$_3^+$ in this high pressure is less sensitive to the change of
  the cross section for X$_2$ + X$_2^+$ collision than in the low pressure.
 This is due to the fact that the increased X$_2$ + X$_2^+$  rate coefficient
  is accompanied by the decreased X$_2^+$ quantity since X$_2$ + X$_2^+$ collision is the main depopulation process for X$_2^+$, which leads to the little change of the X$_3^+$ formation.
  While in the previous case of low pressure, the main depopulating of X$_2^+$
  does not come from the X$_2$ + X$_2^+$ collision but from X$_2^+$ + e
  collision and the density of X$_2^+$ is not affected by the X$_2$ + X$_2^+$  rate coefficient. Thus the density of X$_3^+$ in the low pressure case
  is more sensitive to the change of the cross section for X$_2$ + X$_2^+$
  collision than in the high pressure limit having its linear dependency on the rate coefficient mentioned above. 

  On the other hand, when the molecular temperature is increased from  $T_m$ = 0.026 eV (300 K)  to $T_m$= 4.2 eV (50000 K), the rate coefficient for the present cross section of  H$_2$($v$=0,$j$=0) + H$^2_2$($v'$= 0,$j'$= 0) differs from that for the cross section by Janev et al. \cite{Janev2003} only by about~20 \%. However, the cross section of H$_2$($v$=0,$j$=0) + H$_2^+$($v'$=6,$j'$=0), shown in Figure \ref{Double-fragmentation}, is much larger than that of H$_2$($v$=0,$j$=0) + H$_2^+$($v'$=0,$j'$=0) by Janev et al. \cite{Janev2003} at collision energies higher than ~1 eV. The difference between the rate coefficients using these two cross sections is as much as 10-170 \% at the temperature range $T_m$ = 0.026-4.0 eV. The larger differences are found at higher $T_m$, becoming  over ~40 \% above  1 eV. Hence the high $v'$  state of H$_2^+$ can contribute to the population of H$_3^+$ much more than the $v'$=0 state. This was analyzed 
 by replacing the  H$_2$($v$=0,$j$=0) + H$_2^+$($v'$=0,$j'$=0) reactive rate coefficient
 by that obtained for H$_2$($v$=0,$j$=0) + H$_2^+$($v'$=6,$j'$=0) collision.
  To further analyze the effect of H$_2^+$ vibrational state, new plasma models need to be developed, increasing considerably in complexity for the quantity of processes included.

\begin{table}[h]
    \centering
    \begin{tabular}{llll}\hline
        & Present D$_3^+$ &  Present H$_3^+$ & H$_3^+$ by Janev et al. (2003) \\ \hline\hline
      $\alpha_{16}$ (cm$^{3}$/s) & $2.38 \times 10^{-9}$ & $1.96 \times 10^{-9} [2.76 \times 10^{-9}]$ & $1.59 \times 10^{-9} [2.25 \times 10^{-9}]$ \\
        n$_{D_3^+}$ (cm$^{-3}$) & $1.21 \times 10^{9}$ & $9.95 \times 10^{8} [1.62 \times 10^{9}]$ & $8.12 \times 10^{8} [1.32 \times 10^{9}]$ \\ \hline
    \end{tabular}
    \caption{Rate coefficients for D$_2$($v$=0,$j$=0) + D$_2^+$($v'$=0,$j'$=0) reaction
      and the modeled density of D$_3^{+}$ depending on the cross sections
      shown in Figure \ \ref{D2yD2p-v0-vp0}.a. The values in the $[~~]$ represent for the densities of H$_3^{+}$
      and the rate coefficient for H$_2$($v$=0,$j$=0) + H$_2^+$($v'$=0,$j'$=0) reaction.}
    \label{tab:D3+density}
\end{table}

\section{Conclusion}

In this work a detailed study on the H$_2$ + H$_2^+$($v'$) $\rightarrow$ H$_3^+$ + H reactive cross section
has been done using a quasi-classical treatment,
for collision energies from 1 meV up to 10 eV and for several vibrational
states of the H$_2^+$ reactants and several isotopic variations.
To this aim, new potential energy surfaces have been developed, one to include
non-adiabatic transitions (PESTRIM8$\times$8) and another to increase the accuracy in the whole
energy range up to 17 eV (PES-NN). In all cases, it is found that from 1 meV to $\approx$ 0.5-1 eV
the cross section behaves according to a Langevin law for charge induced-dipole long range interactions.
For energies above 1 eV, the simulated cross section decreases fast,  below the Langevin limit, for all
initial vibrational states of H$_2^+$($v'$). However, for E $>$ 1 eV, the reactive cross section exhibits
a considerable increase with increasing $v'$, and the results seems to converge at $v'$=6.

It is found that the reactive cross section for $v'$=6, summing the H$_3^+$ + H and H$_2$ + H + H$^+$ channels,
match very well the recent experimental measurements by Savic {\it et al.} \cite{Savic-etal:20}, and also
previous measurements \cite{Glenewinkel-Meyer-Gerlich:97,Allmendinger-etal:16b},
describing a broad energy interval from 0.5 meV to 10 eV. Moreover,
the fact that for collision energies above  1 eV the measured reactive cross section show different values in different
experiments, can be explained by a 
 different vibrational excitation of H$_2^+$ achieved in each experimental setup.

Experimentally, the H$_3^+$ products are measured \cite{Savic-etal:20}. The fact that in the QCT simulations, the cross
section for the double fragmentation channel,  H$_2$ + H + H$^+$, needs to be considered to get
an agreement with the new experimental results can be explained by a classical artifact, since
QCT method usually overstimate the energy transfer, specially dealing with systems with equal masses.

The new reactive cross sections obtained for  H$_2$ + H$_2^+$($v'$) and  D$_2$ + D$_2^+$
have been included in a plasma model together with the widely used cross section \cite{Janev2003}.
The resulting population densities of H$_3^+$ and D$_3^+$ are proportional to the rate coefficient,
which in turn indicates the sensitivity of the population density to the adopted cross section.
The new cross sections for vibrational sates ($v'$) of H$_2^+$ will be useful
for the state-resolved CR modeling in plasma with higher molecular temperature,
where higher collision energies (above 0.7 eV) become significant and the differences in
the reactivity 
of the vibrationally excited H$_2^+$ will become important.

\section*{Acknowledgement(s)}

We thank Profs. S. Schlemmer and I. Savi{\'c} for providing us the experimental measurement data
and for very fruitfull discussions. Also, we acknowledge Prof. I. Tanarro for a carefull reading
of the manuscript and very valuable comments and discussions,
and Prof. F. Merkt for providing
their experimental data.  
We  acknowledge computing time at Cibeles (CCC-UAM) and Trueno (CSIC).

\section*{Funding}

The research leading to these results has received fundings from
MICINN (Spain) under grant PID2021-122549NB-C2,
and from KAERI Institutional Program (Korea) under grant No. 524410-22.
Pablo del Mazo-Sevillano acknowledge the finantial support from Ministerio
de Universidades and Universidad Aut\'onoma de Madrid, Plan de Recuperaci\'on, Transformaci\'on y Resiliencia
(Margarita Salas fellowship, Ref. CA4/RSUE/2022-00287).

\appendix
\section{$N$-body permutational invariant polynomials}\label{ap:PIP}
In this section we provide a definition of a permutational invariant polynomial in terms of graph theory, which we latter use as 
a way of filtering the $N$-body polynomials from a general set of PIP.

A graph is a collection of vertices and edges ($\mathcal{G} = (V, E)$) \cite{Trudeau:93}. In an undirected graph the edges are non-ordered pairs of vertices.
An undirected graph is said to be connected if every pair of vertices are joined by a path.

A relation can be established between the polynomials $P_n$ in Eq. \eqref{eq:polynomial_descriptor} and an undirected graph, where the vertices 
represent the particles and the edges the monomials between them.

For instance, the following polynomial
\begin{equation}
    P = p_{12}p_{34}
\end{equation}
can be expressed as the following graph:

\begin{center}
    \begin{tikzpicture}[node distance={15mm}, main/.style = {draw, circle}] 
    \node[main] (1) [above] {$1$};
    \node[main] (2) [right of=1] {$2$}; 
    \node[main] (3) [below of=1] {$3$};
    \node[main] (4) [right of=3] {$4$}; 
    \draw (1) -- (2);
    \draw (3) -- (4);
    \end{tikzpicture} 
\end{center}

This corresponds to a disconnected graph since there are various pairs of vertices which are not reachable, for instance from vertex 2 to 3.
An example of a polynomial whose graph is connected is:
\begin{equation}
    P = p_{12}p_{13}p_{34}
\end{equation}

\begin{center}
    \begin{tikzpicture}[node distance={15mm}, main/.style = {draw, circle}] 
    \node[main] (1) [above] {$1$};
    \node[main] (2) [right of=1] {$2$}; 
    \node[main] (3) [below of=1] {$3$};
    \node[main] (4) [right of=3] {$4$}; 
    \draw (1) -- (2);
    \draw (1) -- (3);
    \draw (3) -- (4);
    \end{tikzpicture} 
\end{center}

A polynomial $P$ is said to be a $N$-body polynomial if its corresponding undirected graph is connected. Any other polynomial that arises as
$\hat{S}P$ with $\hat{S}$ being any operation of a permutation group will be a $N$-body polynomial if $P$ is. Note that the effect of a permutation
operation in the graph only affects the order of the vertices and relabel of the edges:
\begin{equation}
    P' = \hat{S}_{14} P = p_{42}p_{34}p_{31}
\end{equation}

\begin{center}
    \begin{tikzpicture}[node distance={15mm}, main/.style = {draw, circle}] 
    \node[main] (1) [above] {$4$};
    \node[main] (2) [right of=1] {$2$}; 
    \node[main] (3) [below of=1] {$3$};
    \node[main] (4) [right of=3] {$1$}; 
    \draw (1) -- (2);
    \draw (1) -- (3);
    \draw (3) -- (4);
    \end{tikzpicture} 
\end{center}

Given a graph, there are simple algorithms as Depth-First Search (DFS) \cite{Even:11} to compute whether it is a connected graph or not, which recursively
traverses the graph marking each visited vertice. If at the end of the execution all nodes were visited, the graph is connected.
There exist a finite number of connected $N$-vertices undirected graphs which can be evaluated
using the recursive formula \cite{harary:14}:

\begin{equation}
    C_{n}=2^{{n \choose 2}}-{\frac  {1}{n}}\sum _{{k=1}}^{{n-1}}k{n \choose k}2^{{n-k \choose 2}}C_{k}
\end{equation}

These numbers are tabulated for $N$ up to 16 in \cite{mani:16}. One should note that the
number of connected undirected graphs increases fast, for instance for a set of six vertices there exist
26704 of those, so in practice an upper limit on the number of edges has to be set. 

At this 
point we have the tools to determine the number of $N$-body polynomials, as well as, given a polynomial, to determine if it is $N$-body.
If the system presents some kind of permutational symmetry, a minimal set of $N$-body PIP will be generated by projecting the above polynomials 
onto the totally symmetric irreducible representation of the permutation group. The dimension of the minimal PIP set is necessarily lower or 
equal to the minimal one of polynomials, as many of them are related by permutation operations.
Note that we never mentioned the exponents $l$ of the monomials, since they play no role in the graph construction. Hence, what we have defined 
up to here is a generator of $N$-body polynomials or PIPs. Following the general procedure, we are now free to set the desired maximum polynomial 
degree and produce the combinations of monomial exponents $l$ to generate a finite set of functions.

\section{Method for the plasma model\label{ap:plasma}}

The method used to model the plasma is similar to that previously
described \cite{Lishev,Koleva-etal:03}, in which
 each particle's level $i$ density $n_i$ can be solved from a continuity equation   
\begin{equation}
\frac{dn_i}{dt} = \frac{\partial n_i}{\partial t} + \triangledown\cdot(D\triangledown n_i) = \frac{\delta n_i}{\delta t},
\end{equation}
where {\it D} denotes diffusion coefficient and the right hand side is the net particle source
and sink term by collisional-radiative (CR) process and particle flux into and out a volume.
In steady state of $\partial n_i / \partial t = 0$, the density balance equations for a long cylindrical vessel plasma can be expressed as follows.
For atomic levels of H ($n_i , i=1-40$)
\begin{eqnarray}\label{balance}
\frac{dn_i}{dt}&=&\sum_{j>i}^{40} \eta_{ji} A_{ji} n_j - \left(\sum_{j<i}\eta_{ij}A_{ij} + \frac{Q}{V}+\frac{\gamma}{\tau} \delta_{i1}\right)n_i  \nonumber \\
&+& n_e \left(\sum_{j \neq i} \alpha_{1,ji} n_j - \sum_{j \neq i} \alpha_{1,ij} n_i -\alpha_{2i} n_i + \alpha_{41}n_{H^+}\right)  \nonumber \\
&+& n_e \left(\alpha_{5i} + \alpha_6 \delta_{i2} + \alpha_7 \delta_{i1}\right)n_{41} + n_e (\alpha_{9i} + \alpha_{10i} + \alpha_{12i} )n_{42}  \nonumber \\
&+& n_e (\alpha_{13} \delta_{i1} + \alpha_{14} \delta_{i2} + \alpha_{15} \delta_{i1} )n_{43} + n_{41}\alpha_{16} \delta_{i1} n_{42}   \nonumber \\
&+& \left(\sum_{j=42}^{43} \zeta_{aj}\left(\frac{\mu}{R}\right)^2 D_{Aj} n_j + \zeta_{aH^+}\left(\frac{\mu}{R}\right)^2 D_{AH^+} n_{H^+} \right)\delta_{i1} ,
\end{eqnarray}
and for species H$_2 (n_i, i=41)$, H$_2^+ (n_i, i=42)$, and H$_3^+ (n_i, i=43)$ 
\begin{eqnarray}\label{balance1}
\frac{dn_i}{dt}&=& \delta_{i41} \Biggl( n_e \alpha_{14} n_{43} + \frac{Q_{in}}{V}\times4.48\times10^{17}+\frac{\gamma'}{\tau}n_1 
\nonumber \\
& & \quad\quad+\sum_{j=42}^{43} \zeta_{mj} \left( \frac{\mu}{R}\right)^2 D_{Aj} n_j + \zeta_{mH^+}\left(\frac{\mu}{R}\right)^2 D_{AH^+} n_{H^+} \Biggr)  \nonumber \\
&-&n_e \left(\sum_{k=5}^8 \alpha_k \delta_{i41}n_i  +  \sum_{k=9}^{12} \alpha_k \delta_{i42}n_i +\sum_{k=13}^{15} \alpha_k \delta_{i43}n_i \right)   \nonumber \\
&+&\delta_{i42}(n_e \alpha_8 -n_{41}\alpha_{16})n_i +\delta_{i43} n_{41} \alpha_{16} n_{42} -\delta_{i41}\alpha_{16}n_{42}n_i  \nonumber \\
&-&\frac{Q}{V}n_i - (1-\delta_{i41})\left(\frac{\mu}{R}\right)^2 D_{Ai} n_i.
\end{eqnarray}
The density of H$^+$, $n_{H^+}$ is deduced from the quasi neutrality condition
\begin{equation}
n_e = n_{H^+} + n_{42}+n_{43}. 
\end{equation}
{The electron density, $n_e$, is assumed to have a radial distribution close to
  a Bessel-type profile for the ambipolar-diffusion regime considered here and applying the
  Bohm criterion as boundary conditions between the plasma and the vessel walls with
  $\mu$ = 2.405 for an infinite cylinder of the effective radius of $R$
  \cite{Koleva-etal:03,Lieberman-Lichtenberg:05}. The effective $R$ is set as 40 cm for our plasma device.
}

Diffusion time $\tau$ for H atom in the device of radius $R_d$ is given by $\tau = 2R_d/v_{th}$
with the thermal velocity $v_{th} = 2\sqrt{2T_a / \pi M_H}$ for atomic temperature $T_a$ and the hydrogen  mass $M_H$.
The wall recombination coefficient $\gamma$ is given as the empirical expression
$\gamma = 0.151 \exp(-1.09\times10^3 / T_m)$ \cite{Onge}, with $T_m$ being the temperature of H$_2$ molecule.

Under the ambipolar diffusion assumption \cite{Golant} the diffusion coefficient $D_{AH^+}$ for H$^+$ is given by
\begin{equation}
D_{AH^+} = T_e K_1^0 \left( \frac{760}{p}\frac{T_m}{273} \right) ,
\end{equation}
where $p=n_{H_2} T_m$ is in Torr and the electronic temperature $T_e$ is in eV.
The reduced mobility $K_1^0 ({\rm cm}^2{\rm V}^{-1}{\rm s}^{-1}) $
for H$^+$ and D$^+$ are $15.9$ and $11.2$, respectively \cite{McDaniel}.
The diffusion coefficients for H$_2^+$ and H$_3^+$ are given
with the relation $D_{AH^+} : D_{AH_2^+} : D_{AH_2^+} = 1 : \sqrt{2}/\sqrt{3} : \sqrt{5}/3 $ \cite{Onge}.

The conversion coefficients from H ions into H atom ($\zeta_a$)  and H$_2$ molecule  ($\zeta_m$)
on the wall are taken from \cite{Boeuf}. The conversion coefficient $\gamma'$ from H atom to H$_2$ molecule on the wall is also
taken from \cite{Boeuf}. $Q_{in}$, $Q$ and $V$ denote the gas flow rate of H$_2$ in the inlet, the pumping rate in the outlet and the volume of the plasma device, respectively. $Q_{in}$, $Q$ and $V$ are given by 600 sccm (standard cubic centimeters per minute), 4800 lps (liter per second) and 2.64$\times$10$^6$ {\rm cm}$^3$, respectively.

The CR processes and the rate coefficient $\alpha$ notations are listed in Table~\ref{tab:reactions}
with the references of the collision cross section and the radiative transition rate.
The escaping factor $\eta$ for radiation trapping in optically thick plasma is given as that
for the infinite long cylinder \cite{ChaiKwon}.
\begin{table}[h]
  \centering
    \begin{tabular}{lr}
    Reaction & Rate coefficient \\ \hline
    $H(n \geq 1) + e \leftrightarrow H(n' > n) + e$ & $\alpha_1$ \cite{Janev2003}\\
    $H(n \geq 1) + e \leftrightarrow H^+ + 2e$ & $\alpha_2$ \cite{Janev2003}\\
    $H(n \leq 40) \rightarrow H(n' < n) + h\nu$ & $A_{ij}$ \cite{Wiese}\\
    $H^+ + e \rightarrow H(n \leq 40) + h\nu$ & $\alpha_4$ \cite{Janev2003}\\
    $H_2 + e \rightarrow H(n=1) + H(n' \leq 3) + e$ &$\alpha_5$ \cite{Janev1987}\\ 
    $H_2 + e \rightarrow 2H(n=2) + e$ &$\alpha_6$ \cite{Janev1987}\\
    { $H_2 + e \rightarrow H^+ + H(n=1) + 2e$ } & $\alpha_7$ \cite{Janev2003}\\
    $H_2 + e \rightarrow H_2^+ +2e$ & $\alpha_8$ \cite{Janev2003}\\
    {$H_2^+ + e \rightarrow H(n=1) + H(n' \geq 2)$} & $\alpha_9$ \cite{Janev1987} \\
    $H_2^+ + e \rightarrow H^+ + H(n \leq 2) + e$ & $\alpha_{10}$ \cite{Janev1987} \\
    $H_2^+ + e \rightarrow 2H^+ + 2e$ & $\alpha_{11}$ \cite{Janev1987} \\
    $H_2^+ + e \rightarrow H_2^* \rightarrow H(n=1) + H(n' \geq 2)$ & $\alpha_{12}$ \cite{Janev2003}\\
    $H_3^+ + e \rightarrow 3H(n=1)$ & $\alpha_{13}$ \cite{Janev1987} \\
    $H_3^+ + e \rightarrow H_2 + H(n=2)$ & $\alpha_{14}$ \cite{Janev1987} \\ 
    $H_3^+ + e \rightarrow H^+ + 2H(n=1) + e$ & $\alpha_{15}$ \cite{Janev1987} \\
    $H_2 + H_2^+ \rightarrow H_3^+ + H(n=1) $ & $\alpha_{16}$ \cite{Janev2003} \\
    \hline   
    \end{tabular}
    \caption{Collisional-radiative reactions and the rate coefficients considered in our plasma CRM}
    \label{tab:reactions} 
\end{table}

From the cross section  the rate coefficient is obtained as 
\begin{equation}
\alpha = \langle \sigma(v_{12})v_{12} \rangle
\end{equation}
as the averaging over the relative velocity $v_{12}$ distribution.
When the colliding particles of the masses $m_1$ and $m_2$ have the Maxwellian energy distribution
with temperatures $T_1$ and $T_2$ the rate coefficient $\alpha$ can be expressed as \cite{Hagelaar}
\begin{equation}
\alpha(T_{12})=\frac{4}{\sqrt{\pi}v_{T_{12}}^3}\int_0^\infty \sigma(v_{12})\exp(-(v_{12}/v_{T_{12}})^2)v_{12}^3 dv_{12},
\end{equation}
where $T_{12}=(m_2 T_1 + m_1 T_2) / (m_1 + m_2)$ and $v_{T_{12}} = \sqrt{2(m_1 + m_2)T_{12}/m_1 m_2}$
for temperatures in eV. For electron collisions $T_{12}=T_e$
and $v_{T_{12}}=\sqrt{2T_e/m_e}$, with $M_e$ being the electron mass.
The Maxwellian rate coefficient $\alpha_{16}$ for the heavy particle collision
H$_2$ + H$_2^+ \rightarrow$ H$_3^+$ + H($n=1$) is obtained with $T_1 = T_2 = T_m$ and $m_1 = m_2 = 2M_H$.

The balance equations of $dn_i/dt = 0$ including the nonlinear terms
of $\eta_{ij}(n_i)$, $\alpha_{16} n_{41} n_{42}$ and $D_A (n_{41})$ are solved
by the multidimensional secant Broyden’s method \cite{NR} setting the initial $n_i$
as the solution of the linear part of Eqs.\ \ref{balance} and \ref{balance1}
for various $T_e$, $n_e$, $T_m$ and $T_a$.


\end{document}